\documentclass[epj,nopacs]{svjour}
\usepackage{amsfonts}
\usepackage{amsmath}
\usepackage{bm}
\usepackage{etoolbox}
\usepackage{subfig}
\usepackage{graphicx}

\usepackage{tikz}
\usetikzlibrary{arrows,backgrounds,patterns} 
\usetikzlibrary{decorations.pathmorphing,decorations.markings} 
\def\makeheadbox{\relax}

\begin{document}

\renewcommand{\i}{\textrm{i}}       
\newcommand{\dd}{\textrm{d}}        
\newcommand{\mD}{\mathcal{D}}       
\newcommand{\mZ}{\mathfrak{Z}}      
\newcommand{\vv}[1]{\bm{#1}}        
\newcommand{\uv}[1]{\bm{\hat{#1}}}  
\newcommand{\tpsi}{{\widetilde{\psi}}}
\newcommand{\ellp}{\ell_\text{p}}
\newcommand{\etal}{\textit{et~al.}}
\newcommand{\Osize}{0.1}

\newcommand{\psibasic}[1]{
	\begin{scope}[xshift=#1]
		\draw[very thick, black, rotate=15] (0,0) -- (1,0);
		\draw[very thick, black, rotate=-15] (0,0) -- (1,0);
		\draw[very thick, black] (0.966,0.259) arc (105:-105:0.268);
		\draw[fill=black] (0,0) circle (\Osize);
		\draw[thick, ->] (0.4,0) -- (1.1,0);
	\end{scope}
}

\newcommand{\psitildebasic}{
	\begin{scope}
		\draw[very thick, black, rotate=15] (0,0) -- (1,0);
		\draw[very thick, black, rotate=-15] (0,0) -- (1,0);
		\draw[very thick, black] (0.966,0.259) arc (105:-105:0.268);
		\draw[thick, <-] (0.5,0) -- (1.2,0);
	\end{scope}
}

\newcommand{\linebasic}[1]{
	\begin{scope}[xshift=#1]
		\draw[very thick, black,->] (0,0) -- (0.75,0);
		\draw[very thick, black] (0.7,0) -- (1.5,0);
	\end{scope}
}

\newcommand{\crossbasic}{
	\draw[very thick, black] (-0.175,-0.175) -- (0.175,0.175);
	\draw[very thick, black] (-0.175,0.175) -- (0.175,-0.175);
}

\newcommand{\psidiagramO}[1]{
	\begin{tikzpicture}[scale=#1,baseline]
		\psibasic{0.0};
		\draw[fill=black] (0,0) circle (\Osize);
	\end{tikzpicture}
}

\newcommand{\psidiagramangleO}[1]{
	\begin{tikzpicture}[scale=#1,baseline,rotate=45]
		\psibasic{0.0};
		\draw[fill=black] (0,0) circle (\Osize);
	\end{tikzpicture}
}

\newcommand{\psidiagramTILDEO}[1]{
	\begin{tikzpicture}[scale=#1,baseline,rotate=180]
		\psitildebasic;
		\draw[fill=black] (0,0) circle (\Osize);
	\end{tikzpicture}
}

\newcommand{\psidiagramX}[1]{
	\begin{tikzpicture}[scale=#1,baseline]
		\draw[very thick, black, rotate=15] (0,0) -- (1,0);
		\draw[very thick, black, rotate=-15] (0,0) -- (1,0);
		\draw[very thick, black] (0.966,0.259) arc (105:-105:0.268);
		\crossbasic;
		\draw[thick, ->] (0.4,0) -- (1.1,0);
	\end{tikzpicture}
}

\newcommand{\psidiagramTILDEX}[1]{
	\begin{tikzpicture}[scale=#1,baseline,rotate=180]
		\draw[very thick, black, rotate=15] (0,0) -- (1,0);
		\draw[very thick, black, rotate=-15] (0,0) -- (1,0);
		\draw[very thick, black] (0.966,0.259) arc (105:-105:0.268);
		\crossbasic;
		\draw[thick, <-] (0.5,0) -- (1.2,0);
	\end{tikzpicture}
}

\newcommand{\GlineOO}[2]{
	\begin{tikzpicture}[scale=#1,baseline,rotate=#2]
		\linebasic{0.0};
		\draw[fill=black] (0,0) circle (\Osize);
		\draw[fill=black] (1.5,0) circle (\Osize);
	\end{tikzpicture}
}

\newcommand{\GlineXO}[1]{
	\begin{tikzpicture}[scale=#1,baseline]
		\linebasic{0.0};
		\crossbasic;
		\draw[fill=black] (1.5,0) circle (\Osize);
	\end{tikzpicture}
}

\newcommand{\GlineOX}[2]{
	\begin{tikzpicture}[scale=#1,baseline]
		\linebasic{-1.5cm};
		\crossbasic;
		\node[below] at (0,-0.2) {#2};
		\draw[fill=black] (-1.5,0) circle (\Osize);
	\end{tikzpicture}
}

\newcommand{\GlineXX}[3]{
	\begin{tikzpicture}[scale=#1,baseline]
		\linebasic{-1.5cm};
		\crossbasic; 
		\node[below] at (0,-0.2) {#3};
		\begin{scope}[xshift=-1.5cm]
			\crossbasic;
			\node[below] at (0,-0.2) {#2};
		\end{scope}
	\end{tikzpicture}
}


\newcommand{\pfOne}[1]{
	\begin{tikzpicture}[scale=#1,baseline]
		\linebasic{0cm};
		\draw[fill=black] (0,0) circle (\Osize);
		\draw[fill=black] (1.5,0) circle (\Osize);
		\node[below] at (0.3,0) {$z$};
	\end{tikzpicture}
}

\newcommand{\pfTwo}[1]{
	\begin{tikzpicture}[scale=#1,baseline]
		\linebasic{0cm};
		\linebasic{1.5cm};
		\draw[fill=black] (0,0) circle (\Osize);
		\draw[fill=black] (1.5,0) circle (\Osize);
		\draw[fill=black] (3.0,0) circle (\Osize);
		\node[below] at (0.3,0) {$z$};
		\node[above] at (1.5,0.1) {$w$};
		\node[below] at (1.5+0.3,0) {$z$};
	\end{tikzpicture}
}

\newcommand{\pfThree}[1]{
	\begin{tikzpicture}[scale=#1,baseline]
		\linebasic{0cm};
		\linebasic{1.5cm};
		\linebasic{3.0cm};
		\draw[fill=black] (0,0) circle (\Osize);
		\draw[fill=black] (1.5,0) circle (\Osize);
		\draw[fill=black] (3.0,0) circle (\Osize);
		\draw[fill=black] (4.5,0) circle (\Osize);
		\node[below] at (0.3,0) {$z$};
		\node[above] at (1.5,0.1) {$w$};
		\node[below] at (1.5+0.3,0) {$z$};
		\node[above] at (3.0,0.1) {$w$};
		\node[below] at (3.0+0.3,0) {$z$};
	\end{tikzpicture}
}

\newcommand{\pfFour}[1]{
	\begin{tikzpicture}[scale=#1,baseline]
		\linebasic{0cm};
		\linebasic{1.5cm};
		\begin{scope}[xshift=1.5cm,rotate=45]
			\linebasic{0.0cm};
			\draw[fill=black] (1.5,0) circle (\Osize);
		\end{scope}
		\draw[fill=black] (0,0) circle (\Osize);
		\draw[fill=black] (1.5,0) circle (\Osize);
		\draw[fill=black] (3.0,0) circle (\Osize);
		\node[below] at (0.3,0) {$z$};
		\node[above] at (1.5,0.1) {$\xi$};
		\node[below] at (1.5+0.3,0) {$\zeta$};
	\end{tikzpicture}
}

\newcommand{\pfFive}[1]{
	\begin{tikzpicture}[scale=#1,baseline]
		\linebasic{0cm};
		\linebasic{1.5cm};
		\begin{scope}[xshift=1.5cm,rotate=45]
			\linebasic{0.0cm};
			\draw[fill=black] (1.5,0) circle (\Osize);
		\end{scope}
		\draw[fill=black] (0,0) circle (\Osize);
		\draw[fill=black] (1.5,0) circle (\Osize);
		\draw[fill=black] (3.0,0) circle (\Osize);
		\draw[fill=black] (4.5,0) circle (\Osize);
		\node[below] at (0.3,0) {$z$};
		\node[above] at (1.5,0.1) {$\xi$};
		\node[below] at (1.5+0.3,0) {$\zeta$};
		\node[above] at (3.0,0.1) {$w$};
		\node[below] at (3.0+0.3,0) {$z$};
		\linebasic{3.0cm};
	\end{tikzpicture}
}

\newcommand{\pfSix}[1]{
	\begin{tikzpicture}[scale=#1,baseline]
		\linebasic{0cm};
		\linebasic{1.5cm};
		\begin{scope}[xshift=1.5cm,rotate=45]
			\linebasic{0.0cm};
			\draw[fill=black] (1.5,0) circle (\Osize);
		\end{scope}
		\draw[fill=black] (0,0) circle (\Osize);
		\draw[fill=black] (1.5,0) circle (\Osize);
		\draw[fill=black] (3.0,0) circle (\Osize);
		\draw[fill=black] (4.5,0) circle (\Osize);
		\node[below] at (0.3,0) {$z$};
		\node[above] at (1.5,0.1) {$\xi$};
		\node[below] at (1.5+0.3,0) {$\zeta$};
		\node[above] at (3.0,0.1) {$\xi$};
		\node[below] at (3.0+0.3,0) {$\zeta$};
		\linebasic{3.0cm};
		\begin{scope}[xshift=3.0cm,rotate=45]
			\linebasic{0.0cm};
			\draw[fill=black] (1.5,0) circle (\Osize);
		\end{scope}
	\end{tikzpicture}
}

\newcommand{\pfSeven}[1]{
	\begin{tikzpicture}[scale=#1,baseline]
		\linebasic{0cm};
		\linebasic{1.5cm};
		\begin{scope}[xshift=1.5cm,rotate=45]
			\linebasic{0.0cm};
			\draw[fill=black] (1.5,0) circle (\Osize);
		\end{scope}
		\draw[fill=black] (0,0) circle (\Osize);
		\draw[fill=black] (1.5,0) circle (\Osize);
		\draw[fill=black] (3.0,0) circle (\Osize);
		\node[below] at (0.3,0) {$z$};
		\node[above] at (1.5,0.1) {$\xi$};
		\node[below] at (1.5+0.3,0) {$\zeta$};
		\begin{scope}[shift={(2.5607cm,1.0607cm)}]
			\linebasic{0.0cm};
			\draw[fill=black] (1.5,0) circle (\Osize);
			\node[above] at (0.0,0.1) {$\xi$};
			\node[below] at (0.0+0.3,0) {$\zeta$};
			\begin{scope}[rotate=45]
				\linebasic{0.0cm};
				\draw[fill=black] (1.5,0) circle (\Osize);
			\end{scope}
		\end{scope}
	\end{tikzpicture}
}


\newcommand{\psiOne}[1]{
	\begin{tikzpicture}[scale=#1,baseline]
		\linebasic{0cm};
		\crossbasic;
		\draw[fill=black] (1.5,0) circle (\Osize);
	\end{tikzpicture}
}

\newcommand{\psiTwo}[1]{
	\begin{tikzpicture}[scale=#1,baseline]
		\linebasic{0cm};
		\linebasic{1.5cm};
		\crossbasic;
		\draw[fill=black] (1.5,0) circle (\Osize);
		\draw[fill=black] (3.0,0) circle (\Osize);
		\node[above] at (1.5,0.1) {$w$};
		\node[below] at (1.5+0.3,0) {$z$};
	\end{tikzpicture}
}

\newcommand{\psiThree}[1]{
	\begin{tikzpicture}[scale=#1,baseline]
		\linebasic{0cm};
		\linebasic{1.5cm};
		\linebasic{3.0cm};
		\crossbasic;
		\draw[fill=black] (1.5,0) circle (\Osize);
		\draw[fill=black] (3.0,0) circle (\Osize);
		\draw[fill=black] (4.5,0) circle (\Osize);
		\node[above] at (1.5,0.1) {$w$};
		\node[below] at (1.5+0.3,0) {$z$};
		\node[above] at (3.0,0.1) {$w$};
		\node[below] at (3.0+0.3,0) {$z$};
	\end{tikzpicture}
}

\newcommand{\psiFour}[1]{
	\begin{tikzpicture}[scale=#1,baseline]
		\linebasic{0cm};
		\linebasic{1.5cm};
		\begin{scope}[xshift=1.5cm,rotate=45]
			\linebasic{0.0cm};
			\draw[fill=black] (1.5,0) circle (\Osize);
		\end{scope}
		\crossbasic;
		\draw[fill=black] (1.5,0) circle (\Osize);
		\draw[fill=black] (3.0,0) circle (\Osize);
		\node[above] at (1.5,0.1) {$\xi$};
		\node[below] at (1.5+0.3,0) {$\zeta$};
	\end{tikzpicture}
}

\newcommand{\psiFive}[1]{
	\begin{tikzpicture}[scale=#1,baseline]
		\linebasic{0cm};
		\linebasic{1.5cm};
		\begin{scope}[xshift=1.5cm,rotate=45]
			\linebasic{0.0cm};
			\draw[fill=black] (1.5,0) circle (\Osize);
		\end{scope}
		\crossbasic;
		\draw[fill=black] (1.5,0) circle (\Osize);
		\draw[fill=black] (3.0,0) circle (\Osize);
		\draw[fill=black] (4.5,0) circle (\Osize);
		\node[above] at (1.5,0.1) {$\xi$};
		\node[below] at (1.5+0.3,0) {$\zeta$};
		\node[above] at (3.0,0.1) {$w$};
		\node[below] at (3.0+0.3,0) {$z$};
		\linebasic{3.0cm};
	\end{tikzpicture}
}

\newcommand{\psiSix}[1]{
	\begin{tikzpicture}[scale=#1,baseline]
		\linebasic{0cm};
		\linebasic{1.5cm};
		\begin{scope}[xshift=1.5cm,rotate=45]
			\linebasic{0.0cm};
			\draw[fill=black] (1.5,0) circle (\Osize);
		\end{scope}
		\crossbasic;
		\draw[fill=black] (1.5,0) circle (\Osize);
		\draw[fill=black] (3.0,0) circle (\Osize);
		\draw[fill=black] (4.5,0) circle (\Osize);
		\node[above] at (1.5,0.1) {$\xi$};
		\node[below] at (1.5+0.3,0) {$\zeta$};
		\node[above] at (3.0,0.1) {$\xi$};
		\node[below] at (3.0+0.3,0) {$\zeta$};
		\linebasic{3.0cm};
		\begin{scope}[xshift=3.0cm,rotate=45]
			\linebasic{0.0cm};
			\draw[fill=black] (1.5,0) circle (\Osize);
		\end{scope}
	\end{tikzpicture}
}

\newcommand{\psiSeven}[1]{
	\begin{tikzpicture}[scale=#1,baseline]
		\linebasic{0cm};
		\linebasic{1.5cm};
		\begin{scope}[xshift=1.5cm,rotate=45]
			\linebasic{0.0cm};
			\draw[fill=black] (1.5,0) circle (\Osize);
		\end{scope}
		\crossbasic;
		\draw[fill=black] (1.5,0) circle (\Osize);
		\draw[fill=black] (3.0,0) circle (\Osize);
		\node[above] at (1.5,0.1) {$\xi$};
		\node[below] at (1.5+0.3,0) {$\zeta$};
		\begin{scope}[shift={(2.5607cm,1.0607cm)}]
			\linebasic{0.0cm};
			\draw[fill=black] (1.5,0) circle (\Osize);
			\node[above] at (0.0,0.1) {$\xi$};
			\node[below] at (0.0+0.3,0) {$\zeta$};
			\begin{scope}[rotate=45]
				\linebasic{0.0cm};
				\draw[fill=black] (1.5,0) circle (\Osize);
			\end{scope}
		\end{scope}
	\end{tikzpicture}
}


\newcommand{\linepsiX}[1]{
	\begin{tikzpicture}[scale=#1,baseline]
		\linebasic{0cm};
		\psibasic{1.5cm};
		\crossbasic;
		\draw[fill=black] (1.5,0) circle (\Osize);
		\node[above] at (1.5,0.1) {$w$};
		\node[below] at (1.5+0.3,-0.1) {$z$};
	\end{tikzpicture}
}

\newcommand{\linepsipsiX}[1]{
	\begin{tikzpicture}[scale=#1,baseline]
		\linebasic{0cm};
		\psibasic{1.5cm};
		\begin{scope}[xshift=1.5cm,rotate=45]
			\psibasic{0.0cm};
		\end{scope}
		\crossbasic;
		\draw[fill=black] (1.5,0) circle (\Osize);
		\node[above] at (1.5,0.1) {$\xi$};
		\node[below] at (1.5+0.3,-0.1) {$\zeta$};
	\end{tikzpicture}
}


\newcommand{\psiOneTildeX}[1]{
	\begin{tikzpicture}[scale=#1,baseline]
		\linebasic{-1.5cm};
		\crossbasic;
		\draw[fill=black] (-1.5,0) circle (\Osize);
	\end{tikzpicture}
}

\newcommand{\linepsiTildeX}[1]{
	\begin{tikzpicture}[scale=#1,baseline]
		\linebasic{0cm};
		\begin{scope}[rotate=180]
			\psitildebasic{0.0cm};
		\end{scope}
		\begin{scope}[xshift=1.5cm]
			\crossbasic;
		\end{scope}
		\draw[fill=black] (0,0) circle (\Osize);
		\node[above] at (0.0,0.1) {$w$};
		\node[below] at (-0.3,-0.1) {$\widetilde{z}$};
	\end{tikzpicture}
}

\newcommand{\linepsiTildepsiX}[1]{
	\begin{tikzpicture}[scale=#1,baseline]
		\linebasic{0cm};
		\begin{scope}[rotate=180]
			\psitildebasic{0.0cm};
		\end{scope}
		\begin{scope}[rotate=45]
			\psibasic{0.0cm};
		\end{scope}
		\begin{scope}[xshift=1.5cm]
			\crossbasic;
		\end{scope}
		\draw[fill=black] (0,0) circle (\Osize);
		\node[above] at (0.0,0.1) {$\xi$};
		\node[below] at (-0.3,-0.1) {$\widetilde{\zeta}$};
	\end{tikzpicture}
}


\title{Density fields for branching, stiff networks in rigid confining regions}
\titlerunning{Confined branching networks}

\author{Somi\'ealo Azote\inst{1} and Kristian K. M\"uller-Nedebock\inst{1,2}}
\institute{\inst{1} Institute of Theoretical Physics, Department of Physics, Stellenbosch University, Stellenbosch, South Africa\\ \inst{2} National Institute for Theoretical Physics, Stellenbosch, South Africa}


\date{\today}

\abstract{
	We develop a formalism to describe the equilibrium distributions for segments of confined branched networks consisting of stiff filaments. This is applicable to certain situations of cytoskeleton in cells, such as for example actin filaments with branching due to the Arp2/3 complex.
	We develop a grand ensemble formalism that enables the computation of segment density and polarisation profiles within the confines of the cell. This is expressed in terms of the solution to nonlinear integral equations for auxiliary functions. We find three specific classes of behaviour depending on filament length, degree of branching and the ratio of persistence length to the dimensions of the geometry. Our method allows a numerical approach for semi-flexible filaments that are networked.
}

\maketitle


\section{Introduction}
\label{sec:introduction}

Actin filaments forming part of the cytoskeleton of a cell can form branched structures via the Arp2/3 protein complex~\cite{Mullins1998}. This complex nucleates the growth of actin filament daughter branches at an approximate angle of $70 ^{\circ}$ to the predecessor filament~\cite{Small1995,quint2011}. Such branched networks are confined within the cell membrane or by a rigid cell wall in plant cells, which is thought to strongly influence the spatial organisation of these actin networks inside the cell~\cite{Reymann2010}. The abundant actin in cells \cite{dominguez2011actin,Fletcher2010,alberts2002cytoskeleton,lauffenburger1996cell,murzin1995scop} contributes to various processes, including shape remodelling, cell polarity, cell motility or migration or cell crawling for wound healing, cell contractility and division for tissues growth and renewal or regeneration, adhesion, transport  and cell mechanosensing \cite{Fletcher2010,lauffenburger1996cell,Barone2012,Mullins2010,Paluch2009,Kasza2007}. These functions are dependent on correct structure, orientation, and spatial organisation of the actin~\cite{Alvarado2014,Ennomani2016,Michelot2011,Iwasa2007,Hafner2016}.

Actin filaments and microtubules are semi-flexible polymers of which both the persistence lengths and the degrees of polymerization can be similar to the dimensions of the cell within which they are confined~\cite{gittes1993flexural}.  The persistence length of single unconfined actin filaments ranges from $8$ to $25~\mu\text{m}$ and for microtubules is about $5.2~\text{mm}$ \cite{gittes1993flexural,isambert1995flexibility,le2002tracer,ott1993measurement}, while eukaryotic cells can occur in sizes from $10$ to $50~\mu\text{m} $ in animals and up  to $100\,\mu\text{m}$ in plants \cite{Gunning1996}.  Because actin networks are much denser at the cell periphery, the cell size is strongly correlated to cytoskeletal network spatial and orientational organisation, and conformations \cite{Alvarado2014,Ennomani2016,Letort2015}. 

Theoretical and computer simulation studies of single, linear semi-flexible polymers inside spheres~\cite{koster2005brownian,liu2008shapes,milchev2018densely,katzav2006statistical}  show that the mechanical and dynamic properties of the semi-flexible polymers are strongly dependent on both the degree of confinement and the filament persistence length $\ell_{\text{p}}$. (Such insights also help to understand the packaging of very long DNA in bacteriophage capsids \cite{kindt2001dna,odijk2004statics,koster2005brownian}.)
Alvarado~\etal~\cite{Alvarado2014} and Silva~\etal~\cite{Silva2011} have studied the alignment of bundles of semi-flexible filaments inside rectangular chambers with sizes of order of $\mu\text{m}$. They report that spatial confinement has a great impact on the conformational organisation and alignment of actin filaments. This kind of behaviour is also observed for composites of microtubules under confinement~\cite{Tsugawa2016}. Azari \etal~\cite{Azari2015} have also investigated the properties of a mixture of linear polymer chains with stiff segments of different lengths and stiffness in a spherical confining region, using molecular dynamics simulations. In strong confinement, competition in the filament system leads to segregation of chains as the confining volume is decreased.  {Of course, interaction between semi-flexible segments of filaments and their networks can lead to, amongst others, nematic effects. Cross-linking of semi-flexible filaments is known to lead to changes in orientational order~\cite{benetatos:2007}, and there are also nonequilibrium effects arising from active forces (see, \textit{e.g.}, Refs.~\cite{Woodhouse:2012cla,doostmohammadi:2018aa}).}

Here we introduce a theoretical tool with which we can study confined and tree-like branching networks of filaments in equilibrium, suitable for \textit{in vitro} systems and time scales when the many highly out-of-equilibrium cytoskeletal processes in living cells are not applicable.  {In the present paper we restrict ourselves to investigating the interplay of confinement, filament stiffness and tree-like network structure, although the formalism does allow for the inclusion of interactions of various types, but not active forces.} The quantitative theoretical model permits a treatment of the inhomogeneous networks of semi-flexible filaments that emerge in non-infinite regions, which remains challenging to deal with in conventional polymer network theories. We model both the linear and branching networks confined in a spherical domain, where the persistence length $\ell_{\text{p}}$, the contour length $N$, and the sphere diameter $D$ are all of the same order of magnitude. We compare the structural properties of the two types of system. We build upon a formalism for the study of semi-flexible linear polymers confined to a certain region in space from the grand canonical ensemble developed some time ago \cite{Azari2015,Muller2003,Frisch2001} by extending it to branching actin networks. This accounts for filaments with variable lengths and a network where branches can be pruned and regrow. A generalisation of the monomer ensemble technique \cite{Pasquali2009} is particularly suited to computational work, leading to density profiles of branching points, and the distribution of filament orientations which we call the order parameter field in dependence of the proximity to cell boundaries. 

We introduce the grand canonical formalism in Section~\ref{sec:partitionfn} and derive the expressions for the density distributions, the degree of polymerization and the radial order parameter fields for the networks of filament segments. The density functions are expressed in terms of functions that are solutions to a set of non-linear integral equations. We solve these integral equations self-consistently using a recursive numerical scheme. Chemical potentials for linear filament growth and for branching can be varied. Numerical results, presented in Section~\ref{sec:results}, around values where length scales of persistence length, filament degree of polymerization and confining region diameter are more-or-less equivalent, allows a classification of the network behaviour in terms of the local density and order profiles.


\section{The grand partition function of the confined network}
\label{sec:partitionfn}

We calculate the partition function for filament segments that are connected linearly and with branches in a grand ensemble. The resulting equilibrium spatial and orientational distributions of segments can be derived from the solution of integral equations. These allow numerical solutions to be calculated iteratively on appropriate lattices.

For linear chains this has already been done in a similar manner in Refs.~\cite{Muller2003,Frisch2001,Pasquali2009}. We start by summarising the formalism for linear chains in Sec.~\ref{sec:PartFnLinear} in a suitable formalism that we then extend to branching networks in Sec.~\ref{sec:PartFnBranch}.

Our networks consist of two types of building blocks: filament segments and junctions that join segments either linearly or as branches. A filament segment has a specific sense of direction. It starts at a position $\vv{r}_1$, with direction unit vector $\uv{n}_1$, and runs to its other end at position $\vv{r}_1'$ where the  orientation unit vector is $\uv{n}_1'$. Since the filament is polarised, we adopt the convention that $\uv{n}_1$ points away from $\vv{r}_1$, and towards the second end point for $\uv{n}_1'$. The second type of building blocks is the connection between the segment, either connecting them linearly, or as a branching points. The second elements will include a Boltzmann weight to describe a bending energy for the joint, as well as a means of spatially matching the start of one segment to the end of another, etc. 

\subsection{Linear chains in segment ensemble}
\label{sec:PartFnLinear}

We define the statistical weight of a segment of a single oriented filament by $G(\vv{r}_1,\uv{n}_1;\vv{r}_1',\uv{n}_1')=G(\vv{\Gamma}_1)$, where the filament is polarised alongs its contour. The order of the end-points in $G$ therefore matters. Similarly, the fugacity of the filament segment is given by $z(\vv{r}_1,\uv{n}_1;\vv{r}_1',\uv{n}_1')=z(\vv{\Gamma}_1)$, which depends on all the coordinates summarised by $\vv{\Gamma}_1$. 

In a \emph{linear} chain consecutive filament segments are joined head to tail and an additional Boltzmann weight $w(\vv{r}_1',\uv{n}_1';\vv{r}_1\uv{n}_2)$ is required to address the bending energy at each of the joints between the end of the first and the start of the next filament segment. We require the end of one segment and the starting position of the other to be at the same place, so that
\begin{equation}
	w(\vv{r}_1',\uv{n}_1';\vv{r}_1\uv{n}_2) = \delta \left( \vv{r}_1'-\vv{r}_2 \right)\, w_0(\uv{n}_1',\uv{n}_2),
	\label{eq:wDef}
\end{equation}
where $w_0$ is simply a Boltzmann weight associated with the energy penalty for bending the bond.
The grand canonical partition function for the ensemble of filaments joined into a linear chain is then the sum over all filament segment degrees of freedom and numbers of filaments making up a chain.
\begin{align}
	\mathfrak{Z} = & 1+	\int \dd\vv{\Gamma}_1 \,\, z(\vv{\Gamma}_1) \,\, G(\vv{\Gamma}_1) 
	\nonumber \\ & 
	+ \int \dd\vv{\Gamma}_1 \dd\vv{\Gamma}_2 \, z(\vv{\Gamma}_1) \, G(\vv{\Gamma}_1) \, w(\vv{\Gamma}_1, \vv{\Gamma}_2) \, z(\vv{\Gamma}_2) \, G(\vv{\Gamma}_2) + \ldots 
	\nonumber \\ 
	 = & 1+	\int_{1} z_1 \, G_1 +  \int_{1,2} \left( z_1 \, G_1 \right) w_{1,2} \left( z_2 \, G_2 \right) 
	 \nonumber \\
	 & + 
	 \int_{1,2,3} \left( z_1 \, G_1 \right)  w_{1,2} \left( z_2 \, G_2 \right)  w_{2,3} \left( z_3 \, G_3 \right) +
	 \ldots 
	 \label{eq:LinearPartitionFunction}
\end{align}	
We have introduced slightly more compact notation to represent the integration and arguments of $z$ and $G$.

Although by no means necessary, it is practical to introduce filaments of fixed length $\ell$ that are straight and rigid, which means that bends occur only at junctions between segments. This choice means that 
\begin{subequations}
	\begin{align}
	G(\vv{r}_1,\uv{n}_1;\vv{r}'_1,\uv{n}'_1)= & \delta\left( \vv{r}_1' - \vv{r}_1-\ell \uv{n}_1 \right)\delta\left( \uv{n}_1-\uv{n}_1' \right) \text{ and }
	\label{eq:fixedlengthbondG}
	\\
	z(\vv{r}_1,\uv{n}_1;\vv{r}'_1,\uv{n}'_1) = & z(\vv{r}_1,\uv{n}_1)
	\label{eq:fixedlengthbondz}
\end{align}
\end{subequations}
as a simplified fugacity. This form shall be practical to implement on a lattice where the spacing of bonds corresponds to $\ell$ (chosen to be 1) and the possible direction unit vectors are finite.

The spatial confinement of the linear (and also branched) structures is implemented via the segment fugacity:
\begin{subequations}
\label{eqs:constraintimpl}
\begin{align}
	z(\vv{r}_1,\uv{n}_1) =& \begin{cases}
		z_0, & \text{if } \vv{r}_1 \in \mathbb{L} \text{ and } \vv{r}_1+\ell\uv{n}_1 \in \mathbb{L}\\
		0, & \text{otherwise}
	\end{cases}; 
	\label{eq:constraintimpl1}
	\\
	\text{or }\widetilde{z}(\vv{r}_1',\uv{n}_1') =& \begin{cases}
		z_0, & \text{if } \vv{r}_1' \in \mathbb{L} \text{ and } \vv{r}_1'-\ell\uv{n}_1' \in \mathbb{L}\\
		0, & \text{otherwise}
	\end{cases} .
	\label{eq:constraintimpl2}
\end{align}
\end{subequations}
where $\mathbb{L}$ represents the permitted spatial region which the filament segments may occupy. We note that $z$ and $\widetilde{z}$ refer to the fugacity for the same filament, but as expressed in terms of its orientation and position of the start and end points, respectively.

The density of filament segments $\varrho_\text{f}$ starting at position $\vv{r}$ with orientation $\uv{n}$ is then calculated in the usual way from the grand partition function as
 {%
\begin{equation}
	\varrho_\text{f}(\vv{r},\uv{n})  = z(\vv{r},\uv{n}) \frac{\delta\, \ln \mZ[z]}{\delta \, z(\vv{r},\uv{n})}. \label{eq:basic-filament-density}
\end{equation}
}It is the object of this paper to compute and interpret this quantity for the linear and a branched chains in certain confined situations.  {In Section~\ref{sec:PartFnBranch} we shall add the possibility of branching by introducing another fugacity ($\zeta$), allowing a generalisation of eq.~\eqref{eq:basic-filament-density}.} The partition function and the distribution $\varrho_\text{f}$ can be computed for both linear and branched networks in terms of auxiliary functions that are the solutions to integral equations.

\subsubsection{Auxiliary functions for linear chains in a grand ensemble}

For the linear chain we define two functions
\begin{subequations}
\label{eqs:psiLinDefs}
\begin{align}
	 & \psi^\text{(lin)} \left(\vv{r},\uv{n}\right) = 
	\int \dd^3 r_1' \dd^2 n_1' \,\, G(\vv{r},\uv{n}; \vv{r}_1',\uv{n}_1') 
\nonumber \\
	& +
	\int \dd^3 r_1' \dd^2 n_1' \dd \Gamma_2 \,\Bigl[ G(\vv{r},\uv{n}; \vv{r}_1',\uv{n}_1') w(\vv{r}_1',\uv{n}_1';\vv{r}_2,\uv{n}_2) \nonumber \\ & \qquad \times z(\vv{r}_2,\uv{n}_2) \, G(\vv{\Gamma}_2) \Bigr]
 \nonumber \\ &  + 
	\int \dd^3 r_1' \dd^2 n_1' \dd \Gamma_2 \dd \Gamma_3 \,\Bigl[ G(\vv{r},\uv{n}; \vv{r}_1',\uv{n}_1') w(\vv{r}_1',\uv{n}_1';\vv{r}_2,\uv{n}_2) \nonumber \\ & \qquad \times z(\vv{r}_2,\uv{n}_2) \, G(\vv{\Gamma}_2) \, w(\vv{\Gamma}_2,\vv{\Gamma}_3) \nonumber \\ & \qquad \times z(\vv{r}_3,\uv{n}_3) \, G(\vv{\Gamma}_3) \Bigr] + \ldots 
	\label{eq:psiLinDef}
	\\
	& \widetilde{\psi}^\text{(lin)} \left(\vv{r},\uv{n}\right) = 
	\int \dd^3 r_1 \dd^2 n_1 \,\, G(\vv{r}_1,\uv{n}_1; \vv{r},\uv{n}) \nonumber \\
	& +
	\int \dd \Gamma_1 \dd^3 r_2 \dd^2 n_2 \,\Bigl[ G(\vv{\Gamma}_1) \widetilde{z}(\vv{r}_1',\uv{n}_1')  \nonumber \\ & \qquad \times w(\vv{\Gamma}_1,\vv{\Gamma}_2) \, G(\vv{r}_2,\uv{n}_2;\vv{r},\uv{n})\Bigr] 
 \nonumber \\ &  + 
	\int \dd \Gamma_1 \dd \Gamma_2 \dd^3 r_3 \dd^2 n_3 \,\Bigl[ G(\vv{\Gamma}_1) \widetilde{z}(\vv{r}_1',\uv{n}_1') w(\vv{\Gamma}_1,\vv{\Gamma}_2) \nonumber \\ & \qquad \times  G(\vv{r}_2,\uv{n}_2;\vv{r},\uv{n}) \widetilde{z}(\vv{r}_2',\uv{n}_2')  w(\vv{\Gamma}_2,\vv{\Gamma}_3) \nonumber \\ & \qquad \times G(\vv{r}_3,\uv{n}_3;\vv{r},\uv{n})\Bigr] + \ldots 
	\label{eq:psiTildeLinDef}
\end{align}
\end{subequations}
They are simply related to the partition function in eq.~\eqref{eq:LinearPartitionFunction} by an additional integration step
\begin{align}
	\mathfrak{Z} =  & 1 + \int \dd^3 r \dd^2 n \, \, {z}(\vv{r},\uv{n})\, \psi(\vv{r},\uv{n}) 
	\nonumber \\  = &
	1 + \int \dd^3 r \dd^2 n \, \, \widetilde{\psi}(\vv{r},\uv{n}) \, \widetilde{z}(\vv{r},\uv{n}). 
	\label{eq:partition-fn-lin}
\end{align}
Moreover in Appendix~\ref{App:DensityDerivation} we show that the density distribution can be calculated as 
\begin{align}
	\varrho_\text{f}(\vv{r},\uv{n})  = & \frac{z(\vv{r},\uv{n})}{\mathfrak{Z}} \Bigl[ 
		\psi (\vv{r},\uv{n})  \nonumber \\
		& \left. + \int \dd^2 n' \, \widetilde{\psi}(\vv{r},\uv{n}') \, \widetilde{z}(\vv{r},\uv{n}') \, w_0 (\uv{n}',\uv{n}) \, \psi (\vv{r},\uv{n})
	\right]
	\label{eq:filamentdensityfirstinstance}
\end{align}
with $w_0$ as defined in eq.~\eqref{eq:wDef}.

\subsubsection{Diagrammatic representations}

Equations~\eqref{eqs:psiLinDefs} can be written as integral equations for $\psi^{\text{(lin)}}$ and $\widetilde{\psi}^{\text{(lin)}}$, which allows them to be determined numerically by iterative methods. This is clearly seen when we introduce a diagrammatic representation. We denote the filament segment weight by a directed line segment and appropriate labels:
\begin{equation}
	G(\vv{\Gamma}_1) = \delta \left( \vv{r}'_1 - \vv{r}_1 - \ell\uv{n}_1 \right) \delta ( \uv{n}_1-\uv{n}'_1) = \GlineXX{1.0}{$(\vv{r}_1,\uv{n}_1)$}{$(\vv{r}'_1,\uv{n}'_1)$}.
\end{equation}
Integration over all the applicable degrees of freedom at an end is indicated by a filled circle as in
\begin{equation}
	\int \dd^3 r_1\dd^2 n_1\, \delta \left( \vv{r}'_1 - \vv{r}_1 - \ell\uv{n}_1 \right) \delta ( \uv{n}_1-\uv{n}'_1) = \GlineOX{1.0}{$(\vv{r}'_1,\uv{n}_1)$}.
\end{equation}
It follows that the grand partition function can be expressed as 
\begin{align}
	\mathfrak{Z} = & 1+	
	\pfOne{1.0} + \pfTwo{1.0} \nonumber \\ & + \pfThree{1.0} + \ldots
\end{align}
where $z$ below a line indicates appropriate inclusion in the integral of the fugacity function, and the $w$ above a filled circle indicates the Boltzmann weight associated with bending at a junction.
Representing $\psi$ diagrammatically via the diagram \psidiagramX{0.6}, eq.~\eqref{eq:psiLinDef} is represented diagrammatically as
\begin{subequations}
\label{eqs:psiDiag}
	\begin{align}
	\psi^{\text{(lin)}}( \vv{r},\uv{n} )  = &   \psidiagramX{1.0} \nonumber \\
	= & \psiOne{0.7} + \psiTwo{0.7} \nonumber \\ & + \psiThree{0.7} + \ldots \label{eq:psiDiagSelfLong} \\
	= &  \psiOne{0.7} + \linepsiX{0.7}.
	\label{eq:psiDiagSelf}
	\end{align}
\end{subequations}
Eq.~\eqref{eq:psiDiagSelf} can be seen to be equivalent to eq.~\eqref{eq:psiDiagSelfLong} by substituting the LHS into the RHS of the equation recursively, which leads to the infinite sum of the original definition.


\subsection{Branching networks in the filament ensemble}
\label{sec:PartFnBranch}

The main idea of our calculation is to compute the grand partition function $\mZ$ of the system from which one can then obtain the density of filaments and grafts. The grand partition function for a combination of linear bonds and branching junctions has a functional dependence on two fugacities: one for the linear growth of filaments through $z(\vv{r},\uv{n})$, as in the preceding section, and the other through a fugacity $\zeta(\vv{r},\uv{n}_1, \uv{n}_2)$ for a filament grafted in the direction $\uv{n}_2$ from the main filament direction $\uv{n}_1$  at the position $\vv{r}$. In the same way as $w_0$ represents the Boltzmann weight for the bend at the linear junction  {another function} $\xi(\uv{n}'_1,\uv{n}_2,\uv{n}_3)$ represents the Boltzmann weight associated with relative angles the segments make with respect to each other when entering the point where branching occurs.

The grand partition function is the the sum over all possible networks and all conformations of such networks, with appropriate Boltzmann factors and fugacities for linear filaments and branching. It may be expressed diagrammatically as
\begin{align}
	\mathfrak{Z} = & 1+	
	\pfOne{0.6} + \pfTwo{0.6} + \pfThree{0.6} + \ldots
	\nonumber \\
	& + \pfFour{0.6} + \pfFive{0.6} + \ldots \nonumber \\ & + \pfSix{0.6} 
  + \pfSeven{0.6} + \ldots
\end{align}
 {%
For branching networks $\mathfrak{Z}$ is now a functional of both the fugacities $z$ and $\zeta$. Boltzmann factors are shown above the nodes, and the relevant fugacities below the lines.
}

As previously, we define a function $\psi(\vv{r},\uv{n})$ (and its counterpart $\widetilde{\psi}$) such that
\begin{equation}
	\mathfrak{Z} = 1 + \int \dd^3 r\, \dd^2 n\, z(\vv{r},\uv{n}) \, \psi(\vv{r},\uv{n}).
	\label{eq:partfnpsi}
\end{equation}
The function $\psi$ may be expressed diagrammatically, where 
\begin{tikzpicture}[scale=0.6]
	\crossbasic;	
\end{tikzpicture}
indicates that integration over the variables $\{\vv{r},\uv{n}\}$ does not occur,
\begin{align}
	\psi( \vv{r},\uv{n} ) = & \psidiagramX{0.7} 
	\nonumber \\ 
	= &  \psiOne{0.6} + \psiTwo{0.6} + \psiThree{0.6} + \ldots
	\nonumber \\
	& + \psiFour{0.6} + \psiFive{0.6} + \ldots \nonumber \\
	& + \psiSix{0.6}  + \psiSeven{0.6} + \ldots
	\label{eq:psi}
\end{align}
As is familiar from various treatments of tree-like networks, $\psi$ can be written recursively
\begin{subequations}
\label{eq:psiIntegralEq}
	\begin{align}
		\psidiagramX{0.7} = & \,\,\psiOne{0.7} + \linepsiX{0.7} + \linepsipsiX{0.7} \\
		= & \int \dd^3 r_1' \dd^2 n_1' \,\, G(\vv{r},\uv{n}; \vv{r}_1',\uv{n}_1') 
		\nonumber \\ & 
	 + \int \dd^3 r_1' \dd^2 n_1' \dd^2 n_2 \,\Bigl[ G(\vv{r},\uv{n}; \vv{r}_1',\uv{n}_1') w_0(\uv{n}_1',\uv{n}_2) \nonumber \\ & \qquad \times z(\vv{r}_1',\uv{n}_2) \, \psi(\vv{r}_1',\uv{n}_2) \Bigr]
 \nonumber \\ &  +
 \int \dd^3 r_1' \dd^2 n_1' \dd^2 n_2 \dd^2 n_3 \,\Bigl[ G(\vv{r},\uv{n}; \vv{r}_1',\uv{n}_1') 
 \nonumber \\ & \qquad \times \xi (\uv{n}_1',\uv{n}_2,\uv{n}_3) \zeta(\vv{r}_1',\uv{n}_2,\uv{n}_3) \nonumber \\ & \qquad \times
  \psi(\vv{r}_1',\uv{n}_2)\,\psi(\vv{r}_1',\uv{n}_3) \Bigr]
	\end{align}
\end{subequations}
The diagram reresents a nonlinear integral equation, whose solution can then be directly inserted into eq.~\eqref{eq:partfnpsi}. The function $\psi(\vv{r},\uv{n})$ represents the sum of all possible trees with the stem or trunk at $\vv{r}$ in the direction $\uv{n}$.

It is also possible to construct all possible trees from an outer leaf, rather than from the trunk, leading to the definition of  $\widetilde{\psi}(\vv{r},\uv{n})$ in terms of the diagrammatic representation
\begin{align}
	\widetilde{\psi}(\vv{r},\uv{n}) = & \psidiagramTILDEX{0.7} \nonumber \\ = &  \psiOneTildeX{0.7} + \linepsiTildeX{0.7} + \linepsiTildepsiX{0.7}\,\,.
	\label{eq:psitilde}
\end{align}

We can now also compute density distributions for filament segment and branching points:
\begin{subequations}
\begin{align}
	\varrho_\text{f}(\vv{r},\uv{n})  =& z(\vv{r},\uv{n}) \frac{\delta\, \ln \mZ[z,\zeta]}{\delta \, z(\vv{r},\uv{n})} \nonumber \\  =&  \frac{z(\vv{r},\uv{n})}{\mathfrak{Z}} \Bigl[ 
		\psi (\vv{r},\uv{n}) 
		\nonumber \\
		& \left. + \int \dd^2 n' \, \widetilde{\psi}(\vv{r},\uv{n}') \, \widetilde{z}(\vv{r},\uv{n}') \, w_0 (\uv{n}',\uv{n}) \, \psi (\vv{r},\uv{n})
	\right]\label{eq:basic-filament-density1}\\
 \text{ and}\nonumber \\
	\varrho_\text{b}(\vv{r},\uv{n}_1,\uv{n}_2) & = \zeta(\vv{r},\uv{n}_1,\uv{n}_2) \frac{\delta\, \ln \mZ[z,\zeta]}{\delta \, \zeta(\vv{r},\uv{n}_1,\uv{n}_2)}
\nonumber \\ & =\frac{\zeta(\vv{r},\uv{n}_1,\uv{n}_2)}{\mathfrak{Z}} \int \dd^2 n' \,\, \widetilde{\psi}(\vv{r},\uv{n}')\,
\xi (\uv{n}',\uv{n}_1,\uv{n}_2) \nonumber \\ & \qquad \,\psi(\vv{r},\uv{n}_1) \, \psi(\vv{r},\uv{n}_2)
	. \label{eq:basic-branch-density}
\end{align}
\end{subequations}

%


\newpage 

\section{Density distribution and orientational order field profiles}
\label{sec:results}

We modelled the confinement of various type of two dimensional (2d) and three dimensional (3d)  semi-flexible (or stiff) linear chains and branched filament networks in a completely rigid sphere (spherical cell with rigid membrane or wall ) \cite{milchev2018densely}. We  neglect the possibility that the  stiff networks may deform the confining membranes \cite{nakaya2005polymer}.  Analytical derivation of the the local density distributions using the grand partition function (eq.~\eqref{eq:partfnpsi}) led us to the expression in the equations \eqref{eq:psiIntegralEq} and \eqref{eq:psitilde} which are functionals of $\psi$ and $\tilde{\psi}$.

We solve these integral equations  numerically in a  self-consistent manner (Appendix \ref{app:computation} for the numerical method).  {For linear chains and in certain geometries it is possible to provide analytical expressions for the chain density~\cite{Muller2003}, \textit{etc.}. However, eq.~\eqref{eq:partfnpsi} (and its counterpart) become nonlinear when branching is added (the $\zeta$--dependent term). Given the nonlinearity of the equations to solve and wishing to keep the possibility of dealing with arbitrarily-shaped confinement, we present only numerical work on these equations.} We compute the grand canonical partition function and then the density distributions (the average density of linear chain segments, the density of segments involved in branching and the total average density of filaments segments in the confining region) of the system, as well as measures of the order and alignment of the filaments. We express the total density $\varrho$ as a function of the filament bond positions as:
\begin{equation}
\varrho(\pmb{r})=\int \dd^2 n\,\varrho(\pmb{r},\hat{\pmb{n}})
\label{eq:totaldensityr2}
\end{equation}
where $\varrho(\pmb{r},\hat{\pmb{n}})$ is the probability or a density of finding a polymer segment at a position  $\pmb{r}$ and orientation $\hat{\pmb{n}}$ inside the confining region. The density $\varrho(\pmb{r},\hat{\pmb{n}})$ is given by:
\begin{equation}
\varrho(\pmb{r},\hat{\pmb{n}})=\varrho_{\text{f}}(\pmb{r},\hat{\pmb{n}})+\int\dd n_2^2 \, \left[ \varrho_{\text{b}}(\pmb{r},\uv{n},\uv{n}_2) +  \varrho_{\text{b}}(\pmb{r},\uv{n}_2,\uv{n})\right].
\label{eq:totaldensityrn}
\end{equation}
Since we implement our numerical work on a lattice, we continue by summing over both sites and directions rather than integration $\int \dd n^2 \, \ldots \rightarrow \sum_{\uv{n}} \ldots$

For our numerical calculations, we consider a 2d triangular lattice for the growth of 2d branching networks and  3d triangular lattice for the 3d branching networks. On these lattices, we define the spherical confining cells  with diameters $D$ comparable to the persistence length of actin filaments. The typical persistence  length of unconfined actin filament is ($\ell_{\text{p}} \sim 17.7\mu \text{m}$) \cite{gittes1993flexural}. We define the confining region such that the ratio $\ell_{\text{p}}/D \sim 1 .2$ (stiff networks). We express the persistence length of a filament in term of its bending stiffness (see Appendix \ref{app:persistence}).
The polymer chains are allowed to form only inside the confining region with the constraints expressed via eqs.~\eqref{eqs:constraintimpl}. The chain segments 
occupy the bonds (of length $\ell=1$) of the lattice that are within the confining region, otherwise the lattice bonds stay unoccupied. The effective locally oriented linear chain bonds form more readily with the increasing of the fugacity $z_{0}$  while branching is controlled via the fugacity $\zeta_{0}$ at $60^{\circ}$ angle on triangular lattices and as $\zeta_{0}$ increases the network becomes ever more branched. The geometries of the confining regions in our model are controlled by the two fugacities $z_{0}$ and $\zeta_{0}$ which include eqs.~\eqref{eqs:constraintimpl}. The choice of triangular lattices means bonds are multiples of $60^\circ$ which is close to the typically observed $70^\circ$ for Arp2/3-controlled branching.

The mean number of bonds or degree of polymerisation $\langle N \rangle$ of filament segments of a network inside the confining domain is given by
\begin{equation}
\langle N \rangle=\sum_{{\pmb{r}}} \varrho(\pmb{r}).
\label{eq:totaldensityr}
\end{equation}
Therefore the variation of $z_{0}$ and $\zeta_{0}$ allows us to obtain different types of confined networks and the numerical results show distinct differences between the spatial density distribution of filament segment profiles of these networks.

Imaging of actin cytoskeletal networks inside living cells using  electron and fluorescence microscopy has shown highly ordered dense structure in the vicinity of the cell membrane \cite{Alvarado2014,Letort2015,Ennomani2016,fischman1967electron,Atilgan2005,Pritchard2014}. 
Indeed many studies have been conducted to investigate the origin of the ordering of these networks. Starting from the study of  behaviour and conformations of single semi-flexible filament such as actin and DNA under confinement \cite{nakaya2005polymer,liu2008shapes,koster2005brownian,chen2006free,cifra2012weak,dai2014extended}, theoretical studies and coarse-grained computer simulations of semi-flexible cytoskeletal filaments show that confinement of networks of semi-flexible polymers  can induce formation of orientational ordered structure  similar to that observed in molecules of liquid crystalline phase \cite{nikoubashman2017semiflexible,milchev2018densely}. But, to our knowledge, there is no quantitative model that explores the structural spatial organisation and ordering of branching actin networks under cellular confinement in thermodynamic equilibrium. Here we also investigate how confinement affects the alignment of cytoskeletal actin filament networks as their structure and architecture changes from short, long linear filaments to highly branched networks. Although living cells are hardly in equilibrium, we argue that this approach is a sensible first step, that also provides a tool to deal with theoretically challenging finite and confined networks that do not consist of Gaussian chains.

The radial order parameter field $Q(\pmb{r})$ is defined as:
\begin{equation}
Q(\pmb{r})=\sum_{\hat{\pmb{n}}} \frac{1}{2}\left(1-3 (\hat{\pmb{u}}\cdot\hat{\pmb{n}} )^{2})\varrho(\pmb{r},\hat{\pmb{n}}\right)
\label{eqtotalOrderr}
\end{equation}
for 3d networks where $\hat{\pmb{u}}$ is the radial unit vector defined by the positions of the bonds of the filaments in the confining region by $\hat{\pmb{u}}=\pmb{r} / |\pmb{r}|$, with the origin of the coordinate system at the centre of the spherical region.
For 2d networks, $Q(\pmb{r})$ is defined as:
\begin{equation}
Q(\pmb{r})=\sum_{\hat{\pmb{n}}} \left(1-2 (\hat{\pmb{u}}\cdot\hat{\pmb{n}} )^{2})\varrho(\pmb{r},\hat{\pmb{n}}\right).
\label{eqtotalOrderr1}
 \end{equation}
 
For the local radial order field distribution equal to zero, we have a perfectly isotropic distribution of network segments at that point inside the confining region. A positive radial order parameter field value corresponds to alignment of the filaments parallel to the confining cell membrane or wall, whereas a negative radial order parameter field means that the filaments are perpendicular or are pointing straight to the cell membrane.
The profile of the order parameter fields gives us the information about the alignment of the filaments of the networks relatively to the cell membrane.
 
 The density profiles and the order field profile obtained for the networks modelled in 2d are similar to  those modelled in 3d. Depending on the structure and topology of the networks, the confining membrane or wall can have a weak or strong confinement effect on the network organisation. We see from our model that the networks of long linear actin filaments are affected by the strong confinement leading to their bending, while the networks of short filaments are weakly influenced by the effect of confinement \cite{sakaue2007semiflexible,cifra2012weak,chen2006free,liu2008shapes,Smyda2012,Azari2015,Silva2011}. The branched networks, though subjected to a strong confinement, exhibit a high resistance to the effect of confinement.
 
\begin{figure}%
    \centering
    \subfloat[Density profile]{{\includegraphics[width=8.5cm]{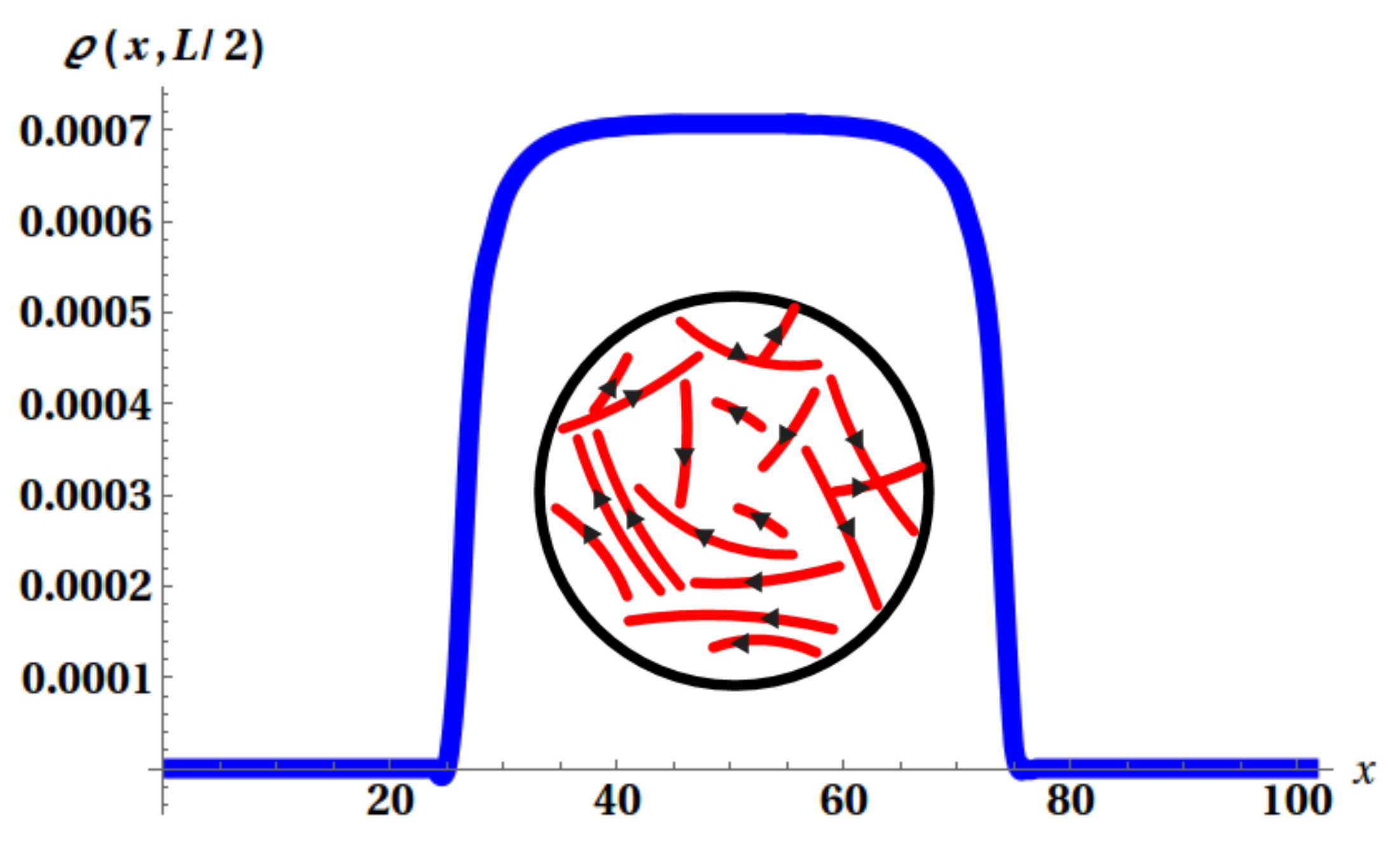} }}%
    \qquad
    \subfloat[Order profile]{{\includegraphics[width=8.5cm]{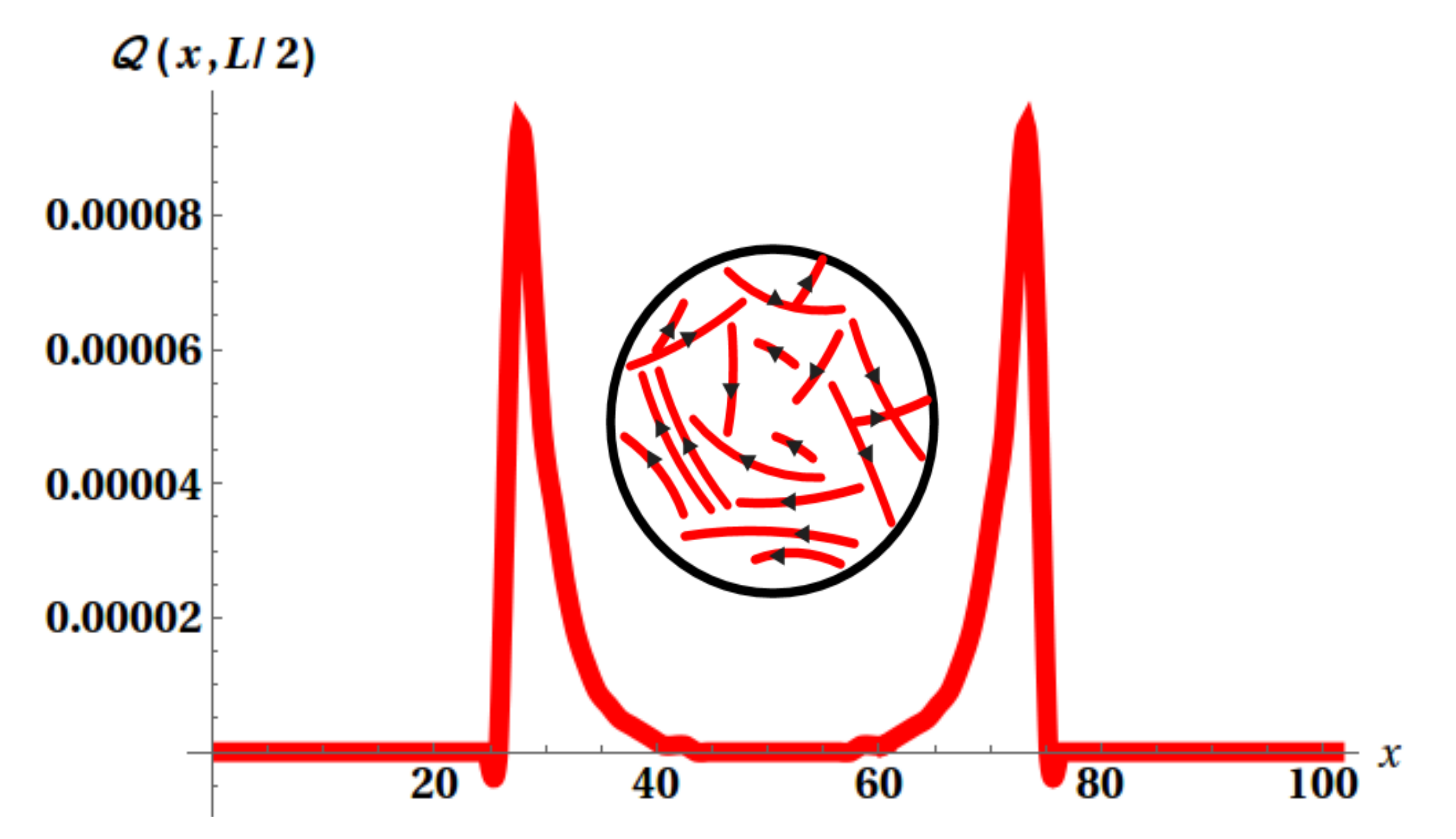} }}%
    \caption{{The average (a) density profile and (b) order parameter field profile through the middle of the sphere (in the direction of $x$) of segments of a 2d network dominated by linear short actin filaments ($\left\langle N \right\rangle<D$), obtained for $z_{0}=0.5$ and $\zeta_{0}=0.001$. The density profile shows a convex-shaped distribution of the network segments inside the confining sphere. The order profile is positive near the cell edges and zero in the middle of the sphere meaning that filaments that are close to the cell membrane are aligned parallel to it, while they are isotropically distributed near the centre of the spherical cell. The inset is a cartoon to illustrate the type of the networks for which the this density profile is obtained.} }%
    \label{fig:5a}%
\end{figure}

\begin{figure}%
    \centering
    \subfloat[Density profile]{{\includegraphics[width=8.5cm]{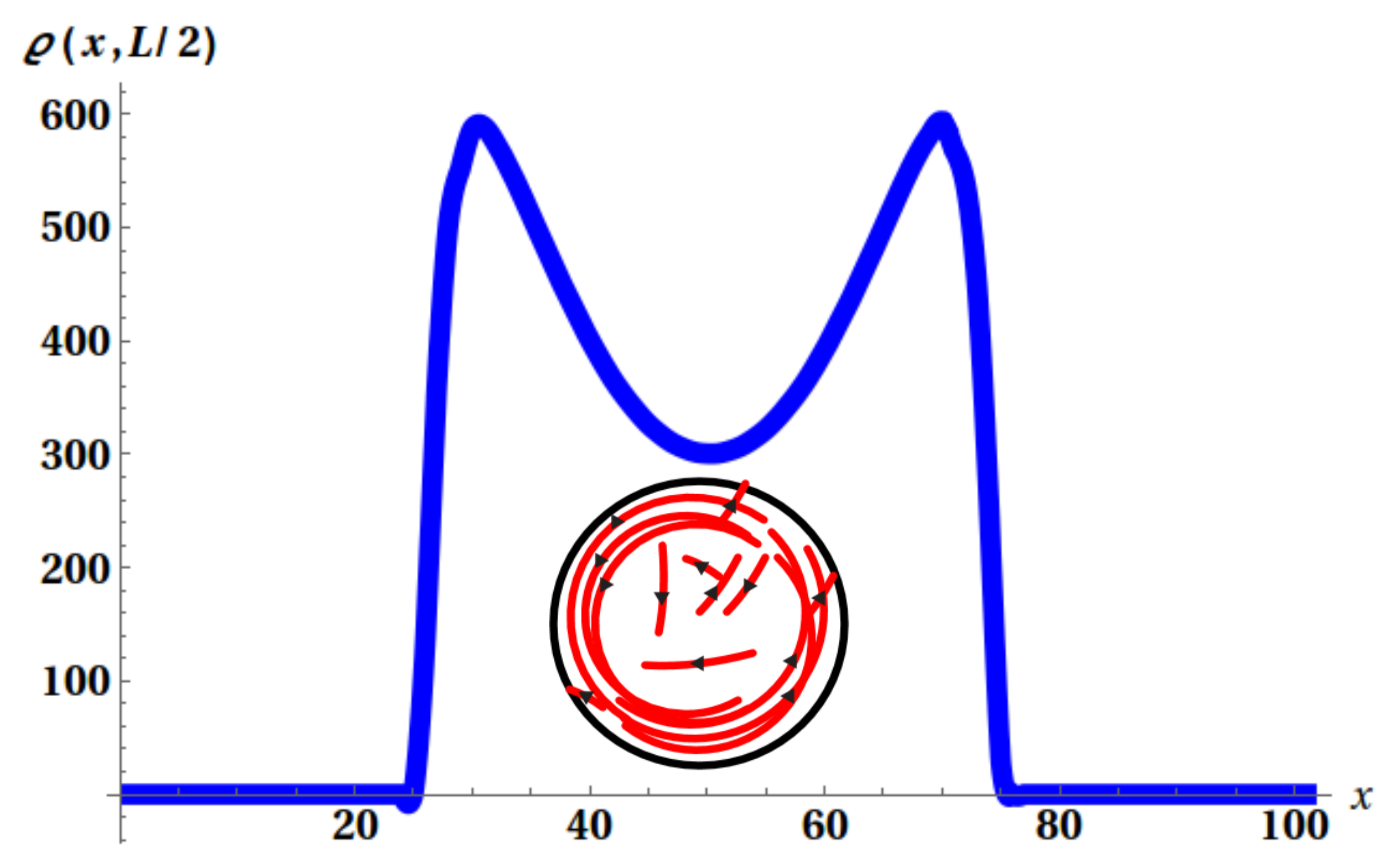} }}%
    \qquad
    \subfloat[Order profile]{{\includegraphics[width=8.5cm]{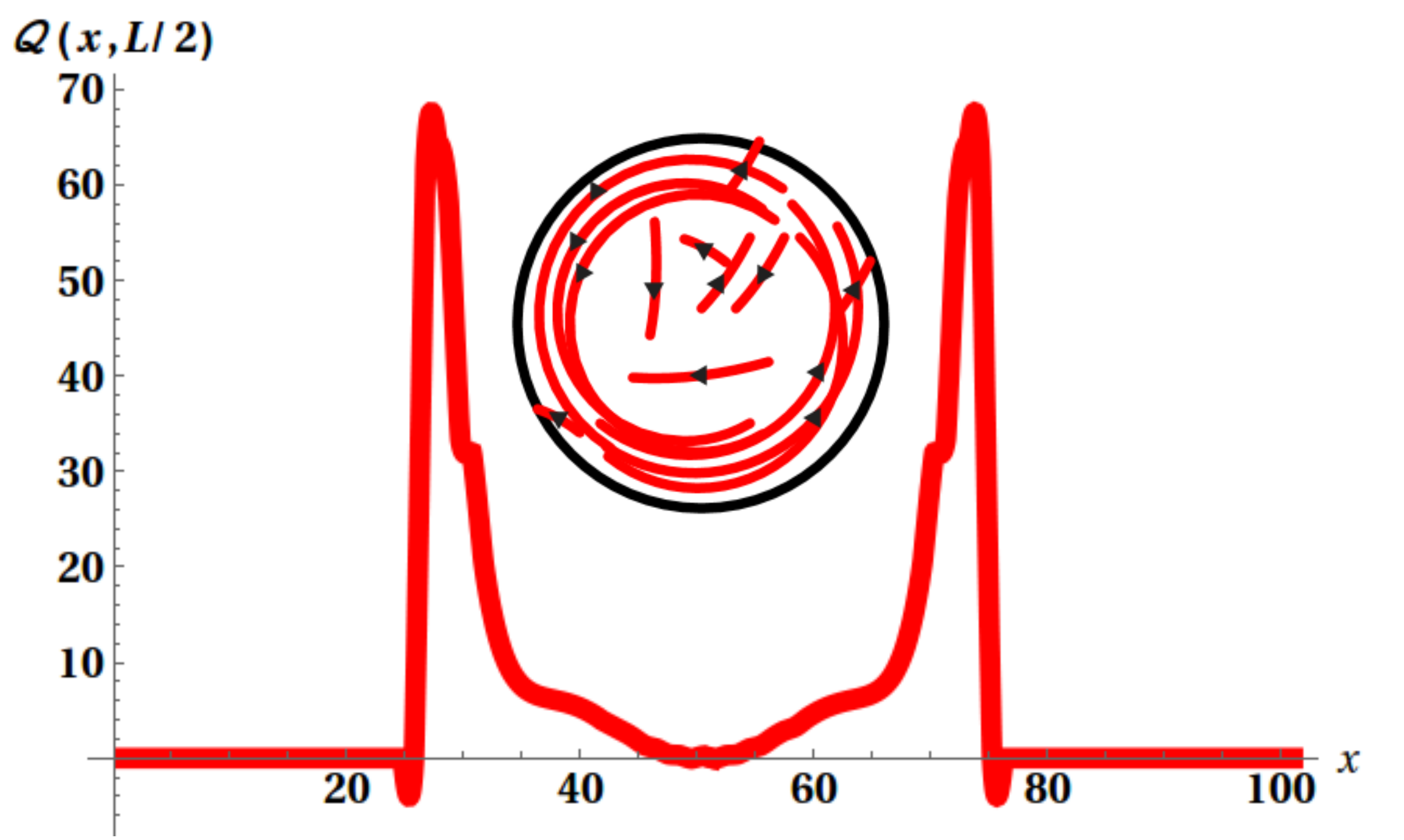} }}%
    \caption{{The average (a) density profile and (b) order parameter field profile through the middle of the sphere (in the direction of $x$) of segments of a 2d network dominated by long linear actin filaments ($\left\langle N \right\rangle> D$) with little branching, obtained for $z_{0}=0.75$ and $\zeta_{0}=0.001$. The density profile shows an inhomogeneous concave-shaped density distribution (high near the periphery of the cell and low near the centre of the sphere) of the filaments or filament segments inside the confining region. The order parameter is positive near the egde of the cell and zero in the middle. This means that the filaments of the network at the cell edges wrap around within the cell and align parallel to it, while those close to the centre are isotropically distributed. The inset is a cartoon to illustrate the type of the networks for which the this density profile is obtained. }}%
    \label{fig:6a}%
\end{figure}

\begin{figure}%
    \centering
    \subfloat[Density profile]{{\includegraphics[width=8.5cm]{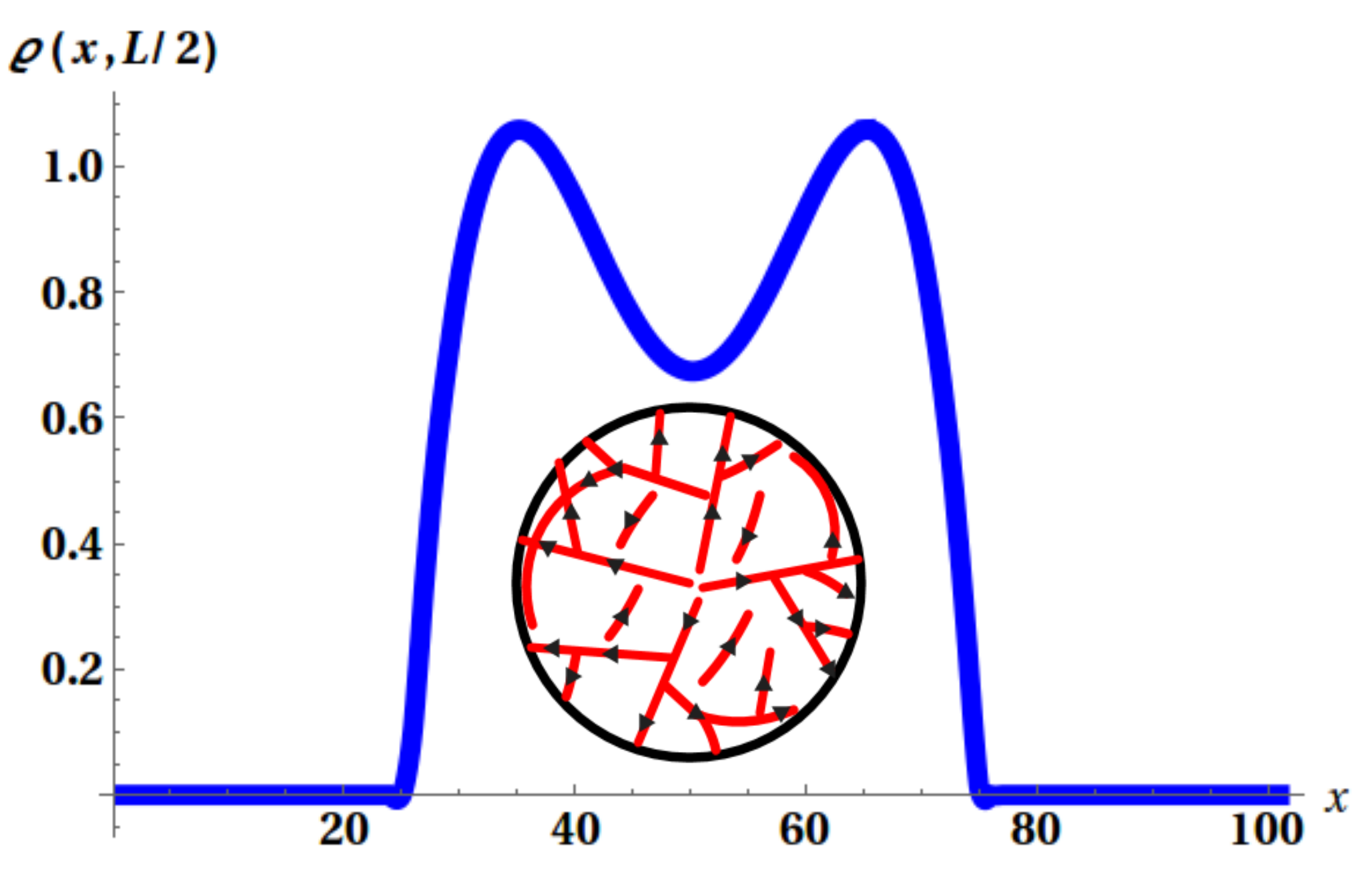} }}%
    \qquad
    \subfloat[Order profile]{{\includegraphics[width=8.5cm]{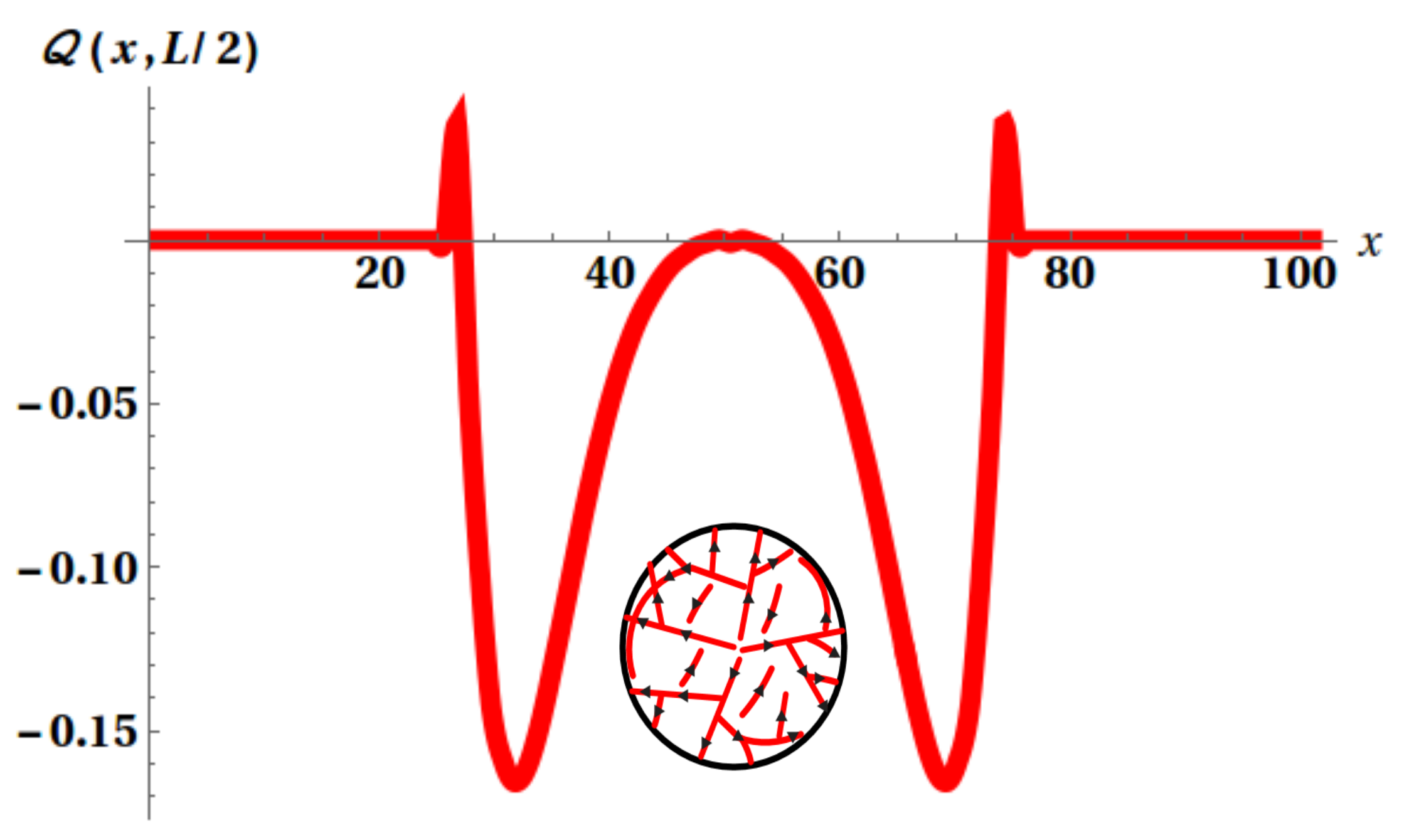} }}%
    \caption{{The graph (a) and graph (b) represent the average density profile and order parameter field profile, respectively, through the middle of the sphere (in the direction of $x$) of segments of a 2d  branched network, obtained for $z_{0}=0.516$ and $\zeta_{0}=0.06$. The density profile shows an inhomogeneous concave-shaped density distributions (high at the cell periphery and low near the centre of the sphere) of the filaments or filament segments inside the confining region.  We obtain a negative order parameter field and this indicates a perpendicular alignment of filaments to the cell membrane or to the cell wall. The inset is a cartoon representing the spherical geometry confining the network of actin filaments in red on the graph and it is there to illustrate the type of the networks for which the this density profile is obtained. }}%
    \label{fig:7a}%
\end{figure}

\begin{figure}%
    \centering
    \subfloat[Density profile]{{\includegraphics[width=8.5cm]{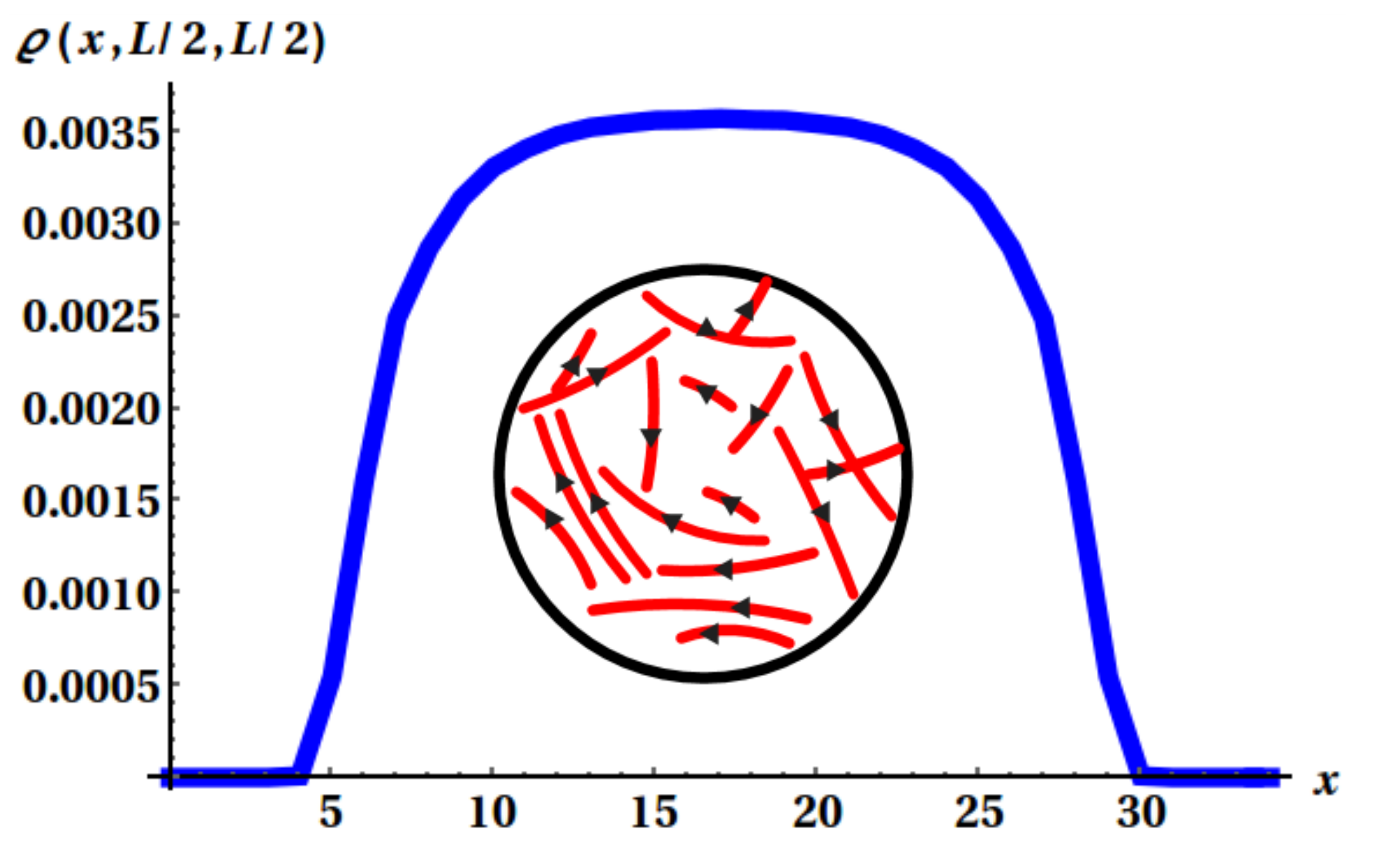} }}%
    \qquad
    \subfloat[Order profile]{{\includegraphics[width=8.5cm]{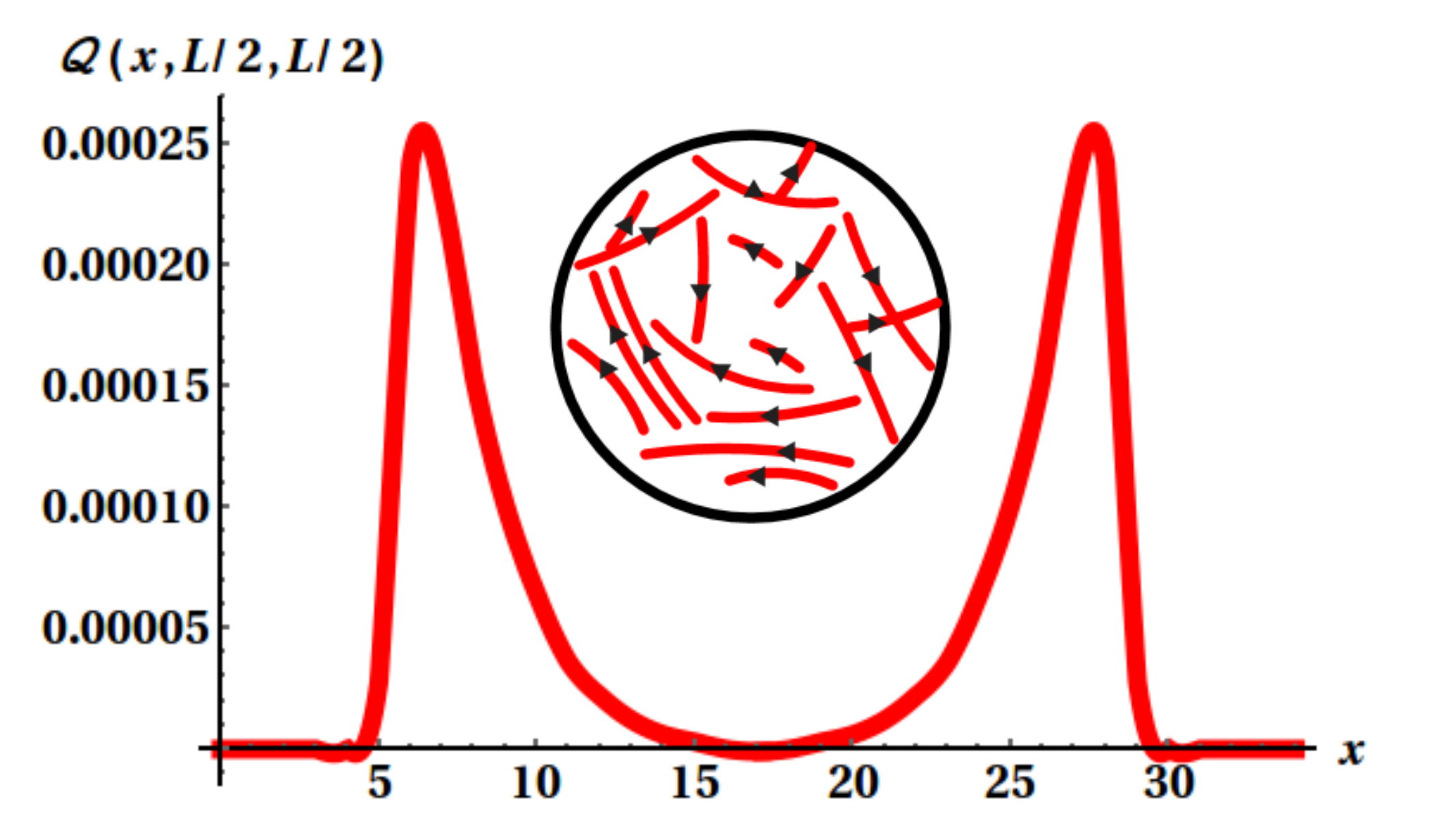} }}%
    \caption{{The plot (a) represents the average density profile and plot (b) the order parameter field profile through the middle of the sphere (in the direction of $x$) of segments of a 3d network dominated by linear short actin filaments ($\left\langle N \right\rangle<D$), obtained for $z_{0}=0.31$ and $\zeta_{0}=0.001$. The density profile shows an convex distribution of the networks segments and relatively flat near the centre of confining spherical cell. The order profile is positive near the cell egde and close to zero in the middle of the sphere meaning that filaments that are close to the cell membrane are aligned parallel to it, while isotropically distributed near the centre of the spherical cell. The inset is a cartoon representing the spherical geometry confining the network of actin filaments in red on the graph and it is there to illustrate the type of the networks for which the this density profile is obtained.} }%
    \label{fig:5b}%
\end{figure}

\begin{figure}%
    \centering
    \subfloat[Density profile]{{\includegraphics[width=8.5cm]{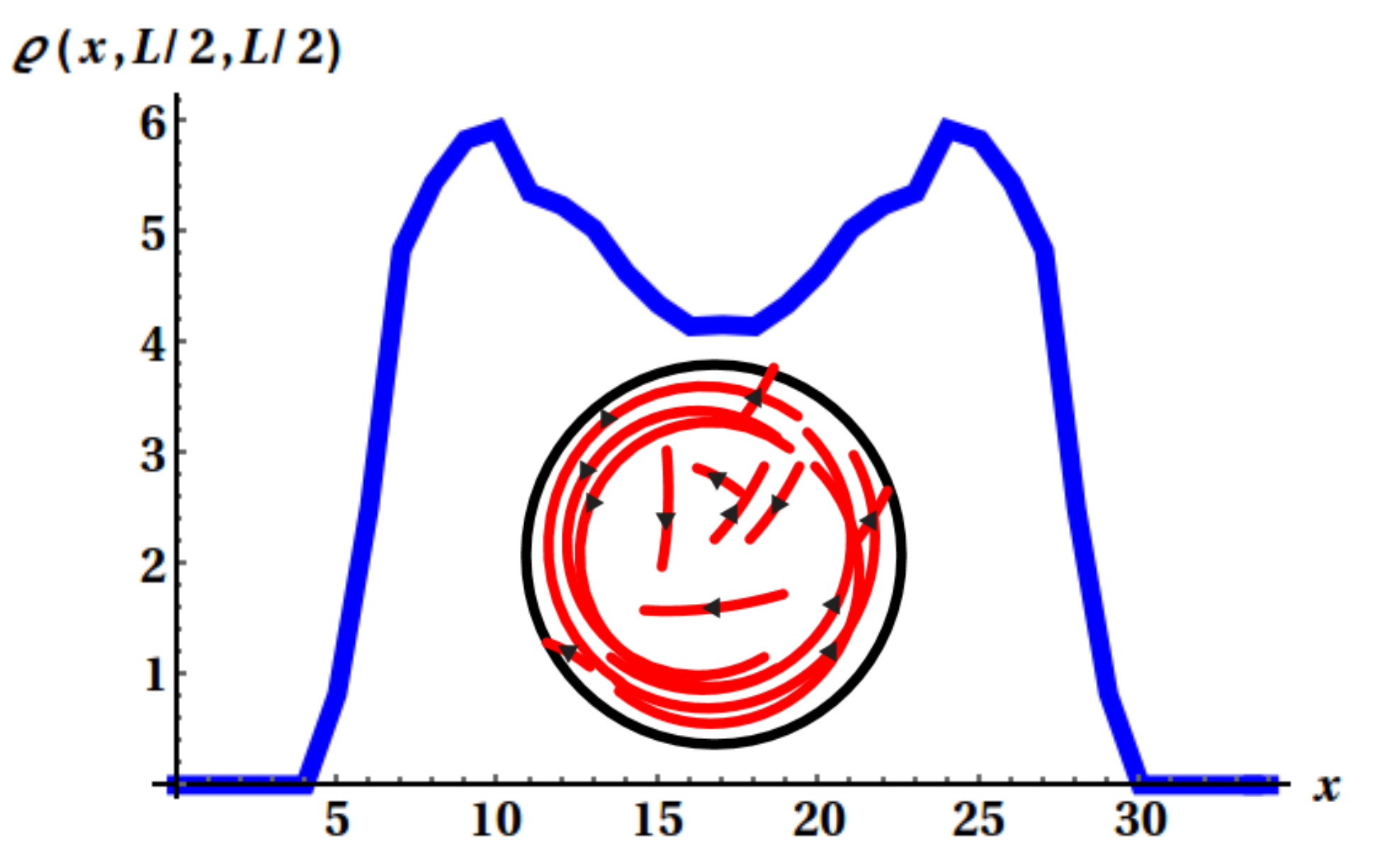} }}%
    \qquad
    \subfloat[Order profile]{{\includegraphics[width=8.5cm]{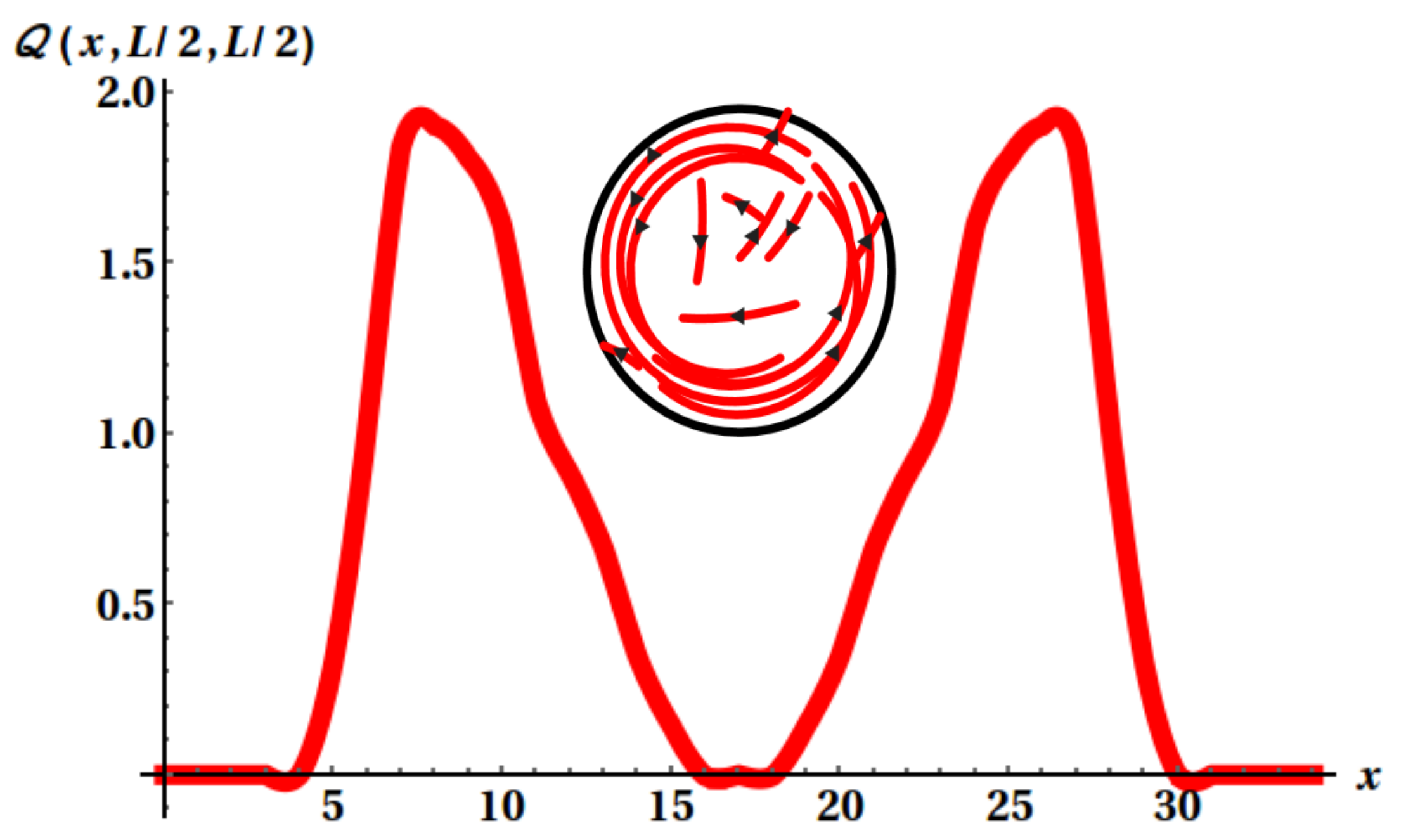} }}%
    \caption{{The graph (a) and graph (b) represent the average density profile and order parameter field profile, respectively, through the middle of the sphere (in the direction of $x$) of segments of a 3d network dominated by long linear actin filaments ($\left\langle N \right\rangle> D$) with little branching, obtained for $z_{0}=0.71$ and $\zeta_{0}=0.00001$. The density profile shows an inhomogeneous concave-like shape density profile with a high segment number at the cell periphery and low near the centre of the spherical cell. The order parameter is positive near the egde of the cell and zero in the middle. This means that the filaments of the networks that are at the cell periphery align parallel to the cell membrane while those close to the centre are isotropically distributed. The inset is a cartoon representing the spherical geometry confining the network of actin filaments in red on the graph and it is there to illustrate the type of the networks for which the this density profile is obtained. }}%
    \label{fig:6b}%
\end{figure}

\begin{figure}%
    \centering
    \subfloat[Density profile]{{\includegraphics[width=8.5cm]{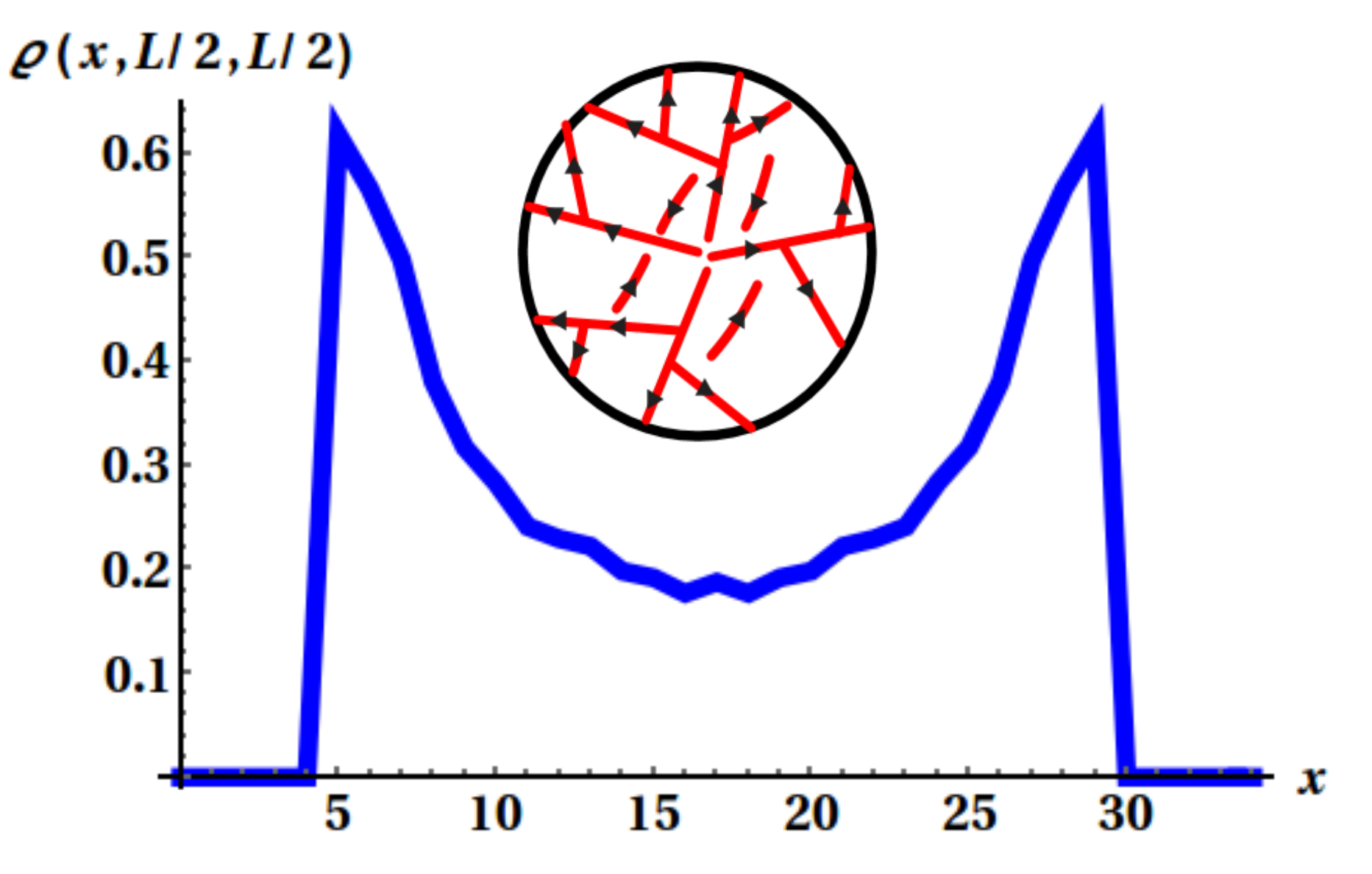} }}%
    \qquad
    \subfloat[Order profile]{{\includegraphics[width=8.5cm]{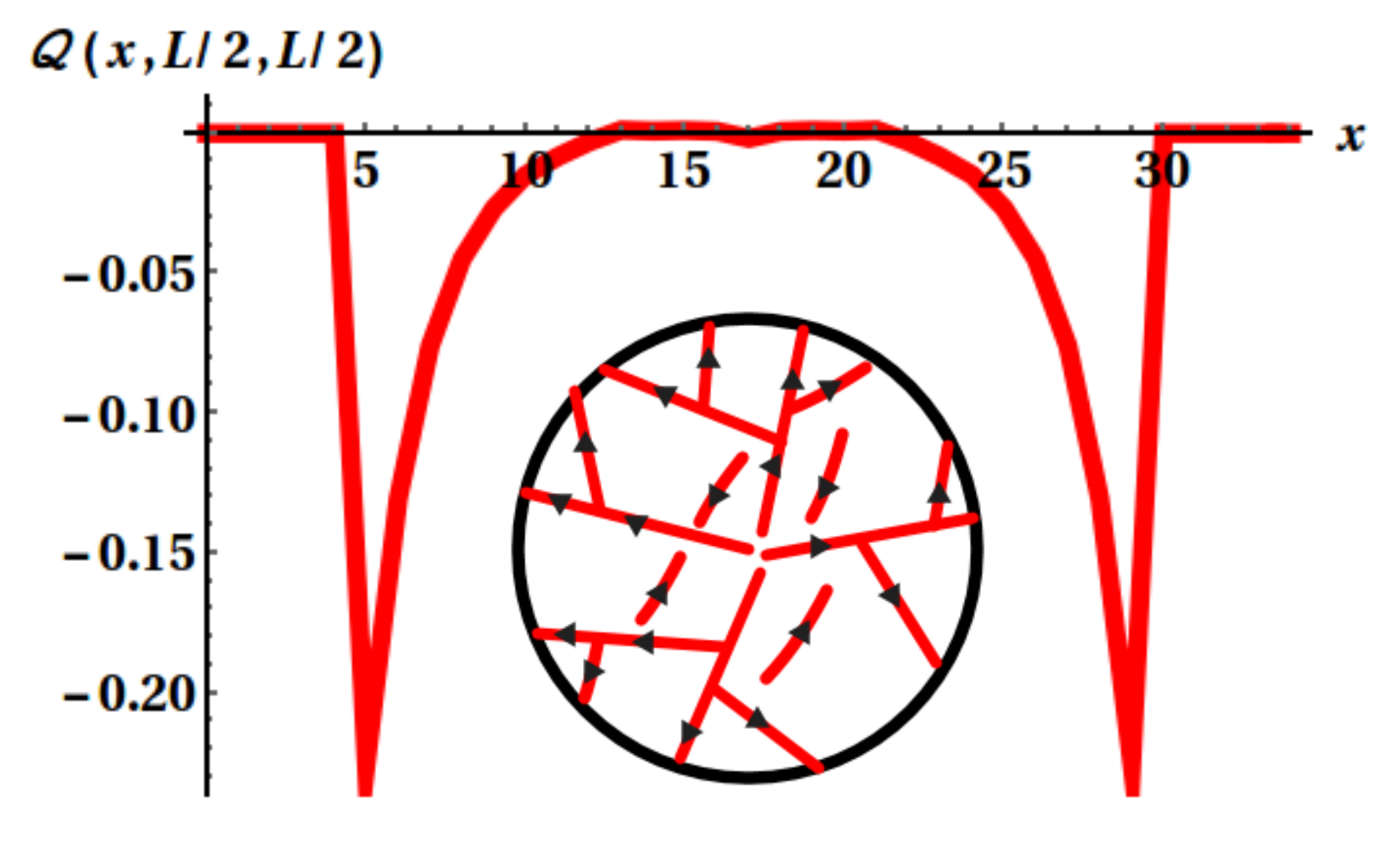} }}%
    \caption{{The graph (a) and graph (b) represent the average density profile and order parameter field profile, respectively, through the middle of the sphere (in the direction of $x$) of segments of a 3d branched network, obtained for $z_{0}=0.5$ and $\zeta_{0}=0.0049$. The density profile shows an inhomogeneous distributions (high at the cell periphery and low near the centre of the sphere) of the filaments or filament segments inside the confining region.  We obtain a negative order parameter field and this indicates a perpendicular alignment of filaments to the cell membrane or to the cell wall. The inset is a cartoon representing the spherical geometry confining the network of actin filaments in red on the graph and it is there to illustrate the type of the networks for which the this density profile is obtained. }}%
    \label{fig:7b}%
\end{figure}


\subsection{Confined actin networks dominated by short filaments ($\left\langle N \right\rangle<D$) }

In our model, we obtain branching actin networks (2d and 3d) dominated by short actin filaments for $\zeta_{0}$ close to zero and small value of $z_{0}$ satisfying the validity domain equation \eqref{eq: condition} such that $(1-z_{0})^{2}> 4\zeta_{0}\sim 0$. We modelled the 2d networks on a the triangular lattice and the 3d networks on a 3d triangular lattice. The structures and properties of the two networks are similar. 
Their density profiles through the middle of the spherical cell (see Figure \ref{fig:5a}(a) for 2d networks and Figure \ref{fig:5b}(a) for 3d networks) show a convex shape density distribution of the filament segments inside the confining domain with a plateau near the centre. This indicates that filaments are concentrated more towards the centre of the sphere. We explain this by the fact that semi-flexible polymers or actin filaments shorter than the confining region size (here the sphere with diameter $D$ compares to the degree of polyemrization $\left\langle N \right\rangle< D$) are weakly influenced by the effects of the confinement and they are free to move toward the centre of the sphere to seek more available volume to orient and translate comfortably and thus maximise their number of accessible conformations. These predictions have been also made in the study of semi-flexible polymer confined in finite regions or between walls \cite{koster2008characterization,nikoubashman2017semiflexible,gao2014free,liu2008shapes}.
The order parameter field profiles in Figures \ref{fig:5a}(b) and Figure \ref{fig:5b}(b) for 2d and 3d networks, respectively, are positive and nonzero near the  wall, becoming zero near the centre. This shows that the short filaments close to the cell membrane or wall have preference of aligning parallel to the cell wall while the filaments close to the centre of the confining domain are isotropically distributed.

\subsection{Confined actin networks dominated by long linear chains ($\left\langle N \right\rangle>D$)}

The actin networks dominated by long linear filaments with very little branching of filaments are obtained for $(1-z_{0})^{2}> 4\zeta_{0}$ with large value of $z_{0}$ while the degree of branching parameter $\zeta_{0}$ is small ($\zeta_{0}$ close to 0).

We plot the the network segment density through the middle of the sphere --- see Figure \ref{fig:6a}(a) for 2d networks and  Figure \ref{fig:6b}(a) for 3d networks. These profiles show a low concentration of filaments near the centre of the sphere while a high concentration is observed close to the edge of the sphere and we qualify this as an heterogeneous concave density distribution of filament segments. 

Previous studies predicted that when long linear semi-flexible polymer filaments are confined in finite geometry with size smaller than their persistence lengths or smaller than their unconfined sizes, they are strongly affected by the presence of the confining membrane. This reduces their available configurations and they bend and occupy the periphery of the confining cell membrane. The density profiles we obtain suggest the same behaviour of the branching actin networks dominated by long linear filaments. So, as the filaments of the networks grow and become longer than the diameter of the sphere,
filaments are subjected to strong confinement effect and they wrap around the spherical cell in order to minimise the free energy of the system \cite{Smyda2012,Azari2015,wang2005generalized,gao2014free,benkova2017structural,chen2016theory}. The order field profiles of these networks through the middle of the sphere (see Figure \ref{fig:6a}(b) for 2d networks and Figure \ref{fig:6b}(b) for 3d networks) confirm these predictions. The order field is greater than zero near the cell membrane while it is close to zero in the middle, indicating that filaments are bent and aligned parallel to the cell membrane while isotropically distributed near the centre of the sphere.

\subsection{Confined branched (actin) Networks}

We obtain branched networks when we increase both $z$ and $\zeta_{0}$ with significant increase of the branching strength $\zeta_{0}$ such that $(1-z_{0})^{2}\simeq 4\zeta_{0}$. A concavely shaped density profile of the network is obtained as we increase $\zeta_{0}$. 
A high number density of branched filament segments is obtained at the cell periphery which is lower near the centre of the cell (inhomogeneous density distribution, as in Figures~\ref{fig:7a}(a) or \ref{fig:7b}(a)). We observe that the branched networks are subjected to a strong confinement effect leading to the inhomogeneous distributions of the networks segments.

The order parameter field profiles through the centre of the sphere of the branched networks are negative on average --- see Figure \ref{fig:7a}(b) for 2d networks and Figure \ref{fig:7b}(b) for 3d networks. This means that most of the filaments of the branched networks align perpendicularly (pointing outwards or inwards) to the cell wall. We expect that this is due to the increase of branching of filaments which makes the network highly stiff and difficult to bend. In fact, some authors have predicted that the increasing of the actin branch nucleation via the Arp2/3 protein complex increases the stiffness of actin networks~\cite{Letort2015,Reymann2010,Iwasa2007}. The effect of rigid strong confinement on branched actin networks is thus cancelled by the stiffness induced by the branching via Arp2/3 protein complex.

With some extensions of our model, one  will be able to elucidate our understanding of at least some equilibrium properties of lamellipodia or filopodia formation inside eukaryotic cells \cite{Borisy2000,Small1995,Lemiere2016}. One can surmise that the stiffness of of branched actin filament network allows it to resist to the effect of confinement that the cell membrane introduce and leads to the protrusion of the membrane if the latter is semi-flexible \cite{nakaya2005polymer}. 
 
\subsection{Types of behaviour} 

\begin{figure}[!h]
\centering
\includegraphics[width=0.5\textwidth]{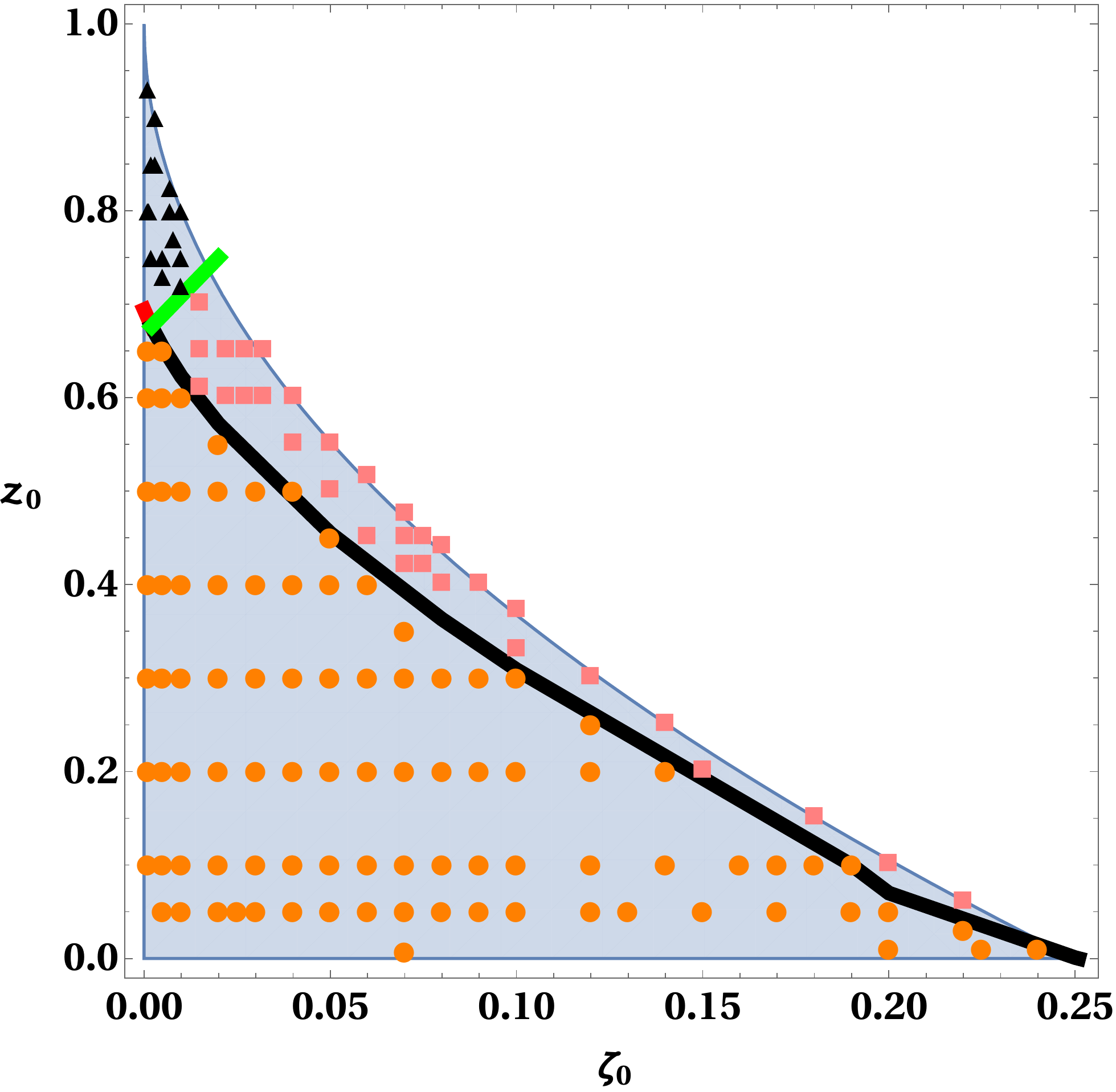}
\caption{{Categories of behaviour for 2d simulation showing the domains of the three studied type of networks. The number densities of filament segments are plotted as we vary the filament elongation parameter $z_{0}$  and the branching parameter $\zeta_{0}$. On this phase diagram, the region represented by yellow filled circles corresponds to the domain in which networks of stiff short filaments with few branching is obtained. The black triangle domain is mainly the zone of networks dominated by long linear chains while little branching is observed. The red squares domain corresponds to the domain of branched actin networks. The points of the line separating the different domain correspond to the points ($\zeta_{0},z_{0}$) for which we obtain the networks with comparable length scales (\textit{i.e.} $\left\langle N \right\rangle \sim D \sim \ell_{\text{p}} $).
The whole blue domain is the domain in which our numerical result is valid (validity domain) and this is given by the equation} (\ref{eq: condition}). Please see Table~\ref{table:criteria}.}
\label{fig:Figp1}
\end{figure}

\begin{figure}[!h]
\centering
\includegraphics[width=0.5\textwidth]{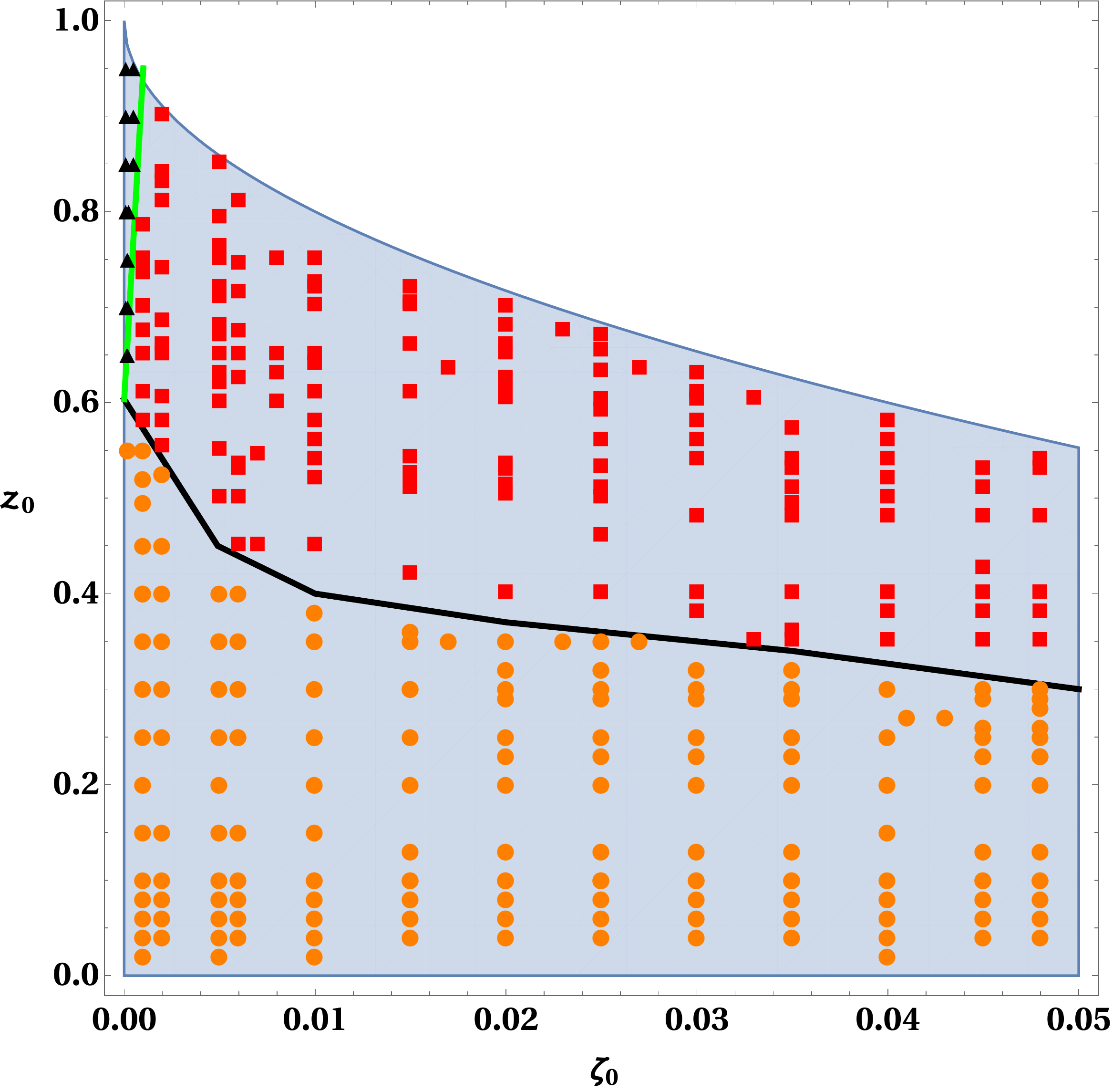}
\caption{{Classification of behaviour in 3d for triangular lattice simulation. The regions of the graphs are distinguished in the same manner as for the 2d triangular lattice, Fig.~\ref{fig:Figp1}. Please also Table~\ref{table:criteria} for clarifications of the different regions.}}
\label{fig:Figp2}
\end{figure}

We have classified the differents networks that form in phase diagrams $(\zeta_{0},z_{0})$, given in   Figure \ref{fig:Figp1} and Figure \ref{fig:Figp2} for 2d  and 3d networks, respectively.

We have computed the densities and the order parameters fields for many points $(\zeta_{0},z_{0})$ that satisfy the domain of validity condition of our model given by the equation (\ref{eq: condition}). The profiles of these densities and order fields allowed us to divide the validity domain into  three distinct zones which corresponds to the three different types of networks we presented in the above.
The yellow dots zone on the phase diagram corresponds to points $(\zeta_{0},z_{0})$ for which networks dominated by short linear filaments are formed. They are are more concentrated near the centre of the confining cell, weakly influenced by the confiment effect with parallel alignment to the cell wall or membrane preference.
 \begin{table}[!h]
 \begin{center}
  \begin{tabular}{ | c | c | c |}
     \hline
\emph{Curvature at centre} & \emph{Av.~order param.} & \emph{Symbol} \\
  \hline 
 $>0$ & $>0$  & black triangles    \\
  \hline
 $>0$ & $ <0$ & red square \\
  \hline 
  $<0$ & $>0$ & yellow disks   \\
  \hline
  \end{tabular}
  \end{center}
  \caption{Summary of the description of the regions in Figs.~\ref{fig:Figp1} and \ref{fig:Figp2}.}
  \label{table:criteria}
  \end{table}
The properties of the networks that are represented in the black triangle domain are those of the branching actin networks dominated by long linear actin filaments with little branching.  They have a positive order parameter suggesting that the filaments are aligned parallel to the cell wall. 
The points $(\zeta_{0},z_{0})$ in the red squares zone on the phase diagram satisfy $(1-z_{0})^{2}\simeq 4\zeta_{0}$. This zone corresponds to confined branched networks filament segments with negative order parameter fields meaning that the filaments of the networks are aligned perpendicular to the cell wall.

\begin{figure}%
    \centering
    \subfloat[Density profile]{{\includegraphics[width=8.5cm]{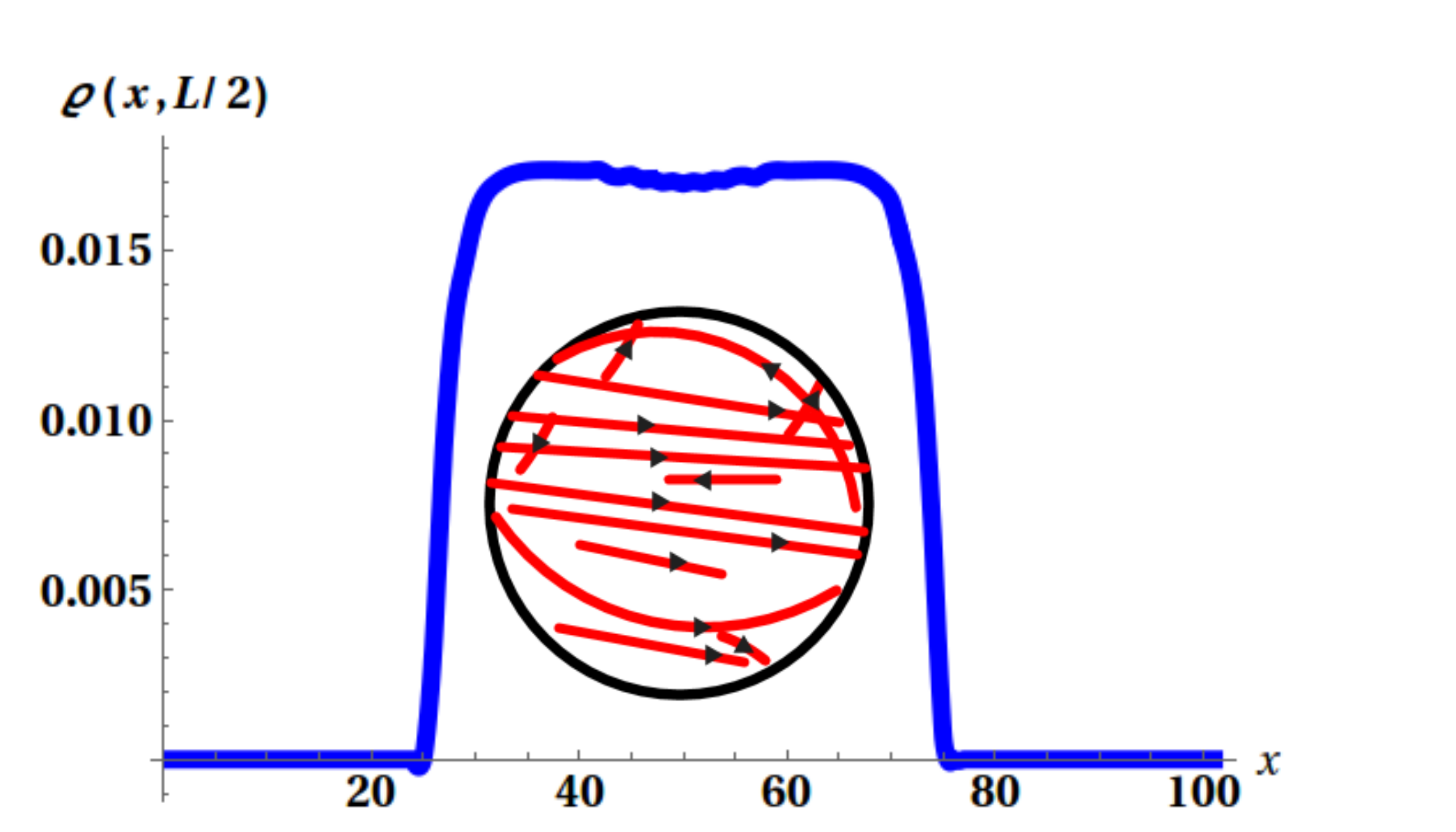} }}%
    \qquad
    \subfloat[Order profile]{{\includegraphics[width=8.5cm]{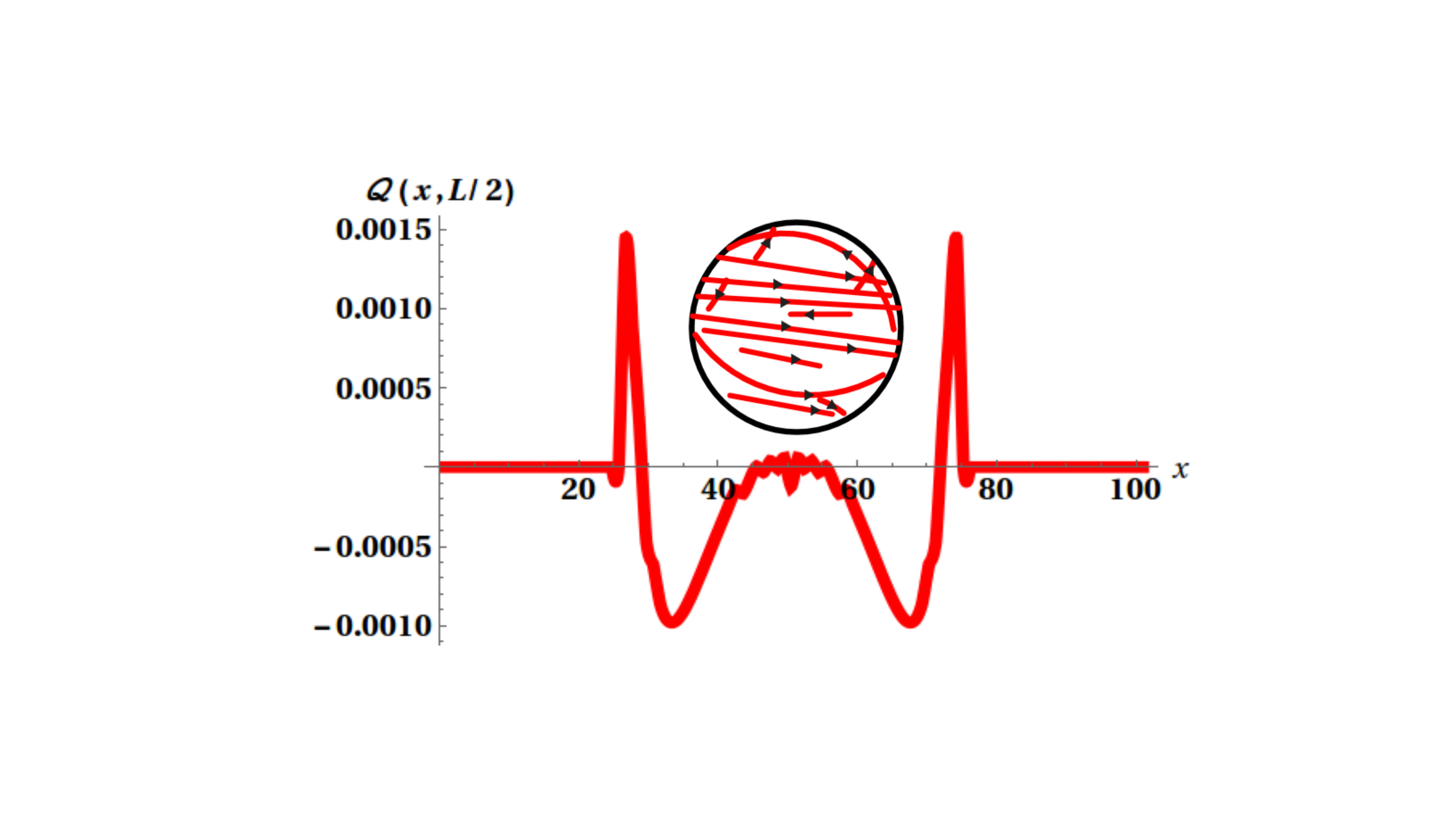} }}%
    \caption{{The graph (a) and graph (b) represent the average density profile and order parameter field profile through the middle of the sphere (in the direction of $x$) of segments of a network in which the length of the filaments are equal to the size of the confining region ($ \left\langle N \right\rangle \sim D \sim \ell_{\text{p}} $). We plot this density profile for $\zeta_{0}=0.001$ and $z_{0}=0.69$. The small dip in the middle of the density profile suggess a decrease of segment density distribution in the centre of the sphere. The inset is a cartoon representing the spherical geometry confining the network of actin filaments in red on the graph and it is there to illustrate the type of the networks for which the this density profile is obtained. }}%
    \label{fig:8b}%
\end{figure}

 In the region of points ($\zeta_{0},z_{0}$) for which we obtain the networks with comparable length scales (\textit{i.e.}. $\left\langle N \right\rangle \sim D \sim \ell_{\text{p}} $) (see the density and order profiles of these networks on Figure \ref{fig:8b} where some of the filaments start bending while others are radial, \textit{i.e.}.~perpendicularly to the confining cell wall or membrane) are used to draw the lines separating the domains of the three types of networks.

%

\section{Conclusion}
\label{sec:Conclusion}

In this work, we have extended the monomer ensemble formalism \cite{Muller2003,Frisch2001,Azari2015} to investigate some of the structural and physical properties of branching actin cytoskeletal networks in spherical rigid confining regions.
The wall, or confinement was treated purely as such, without any additional interactions. Our calculations in thermodynamic equilibrium neglected the excluded volume effect between actin monomers, the interactions between chains and those between chains and the cell membrane or wall, but the inclusion of these interactions is certainly possible in the formalism. By varying the fugacity $z_{0}$  associated to the degree of polymerization of a filament, the fugacity $\zeta_{0}$ associated to the degree of branching, and the ratio of the persistence length and the confining regions diameter, we have identified three distinct types of branching actin networks by their computed density and order parameter field profiles.
  
  {We note that here we have chosen to isolate the role of confinement together with the inherent connectivity of the stiff networks in investigating how they influence filament density and orientation. Certainly, hard-core mutual repulsion of filament monomers, nematic interactions, cross-linking~\cite{benetatos:2007}, and the non-equilibrium action of molecular motors (see Refs.~\cite{Woodhouse:2012cla,doostmohammadi:2018aa}) also are known to affect orientation and density. Indeed, the grand canonical formalism can cope with interaction mechanisms arising from potential energies of interaction \cite{Pasquali2009,azote:2018thesis}, in which there is an interplay of more effects and associated length scales. Furthermore, even a hard core repulsion between infinitely thin linear objects is expected to exhibit different ordering behaviour in three or two dimensions (see, amongst many others, Ref.~\cite{eppenga:1985,vink:2014,mederos:2014}). In this context we remark that most (simple) instances of the grand canonical formalism would only provide a mean-field density functional perspective on such phenomena.
 } 
  
When (statistically) short filaments ($\left\langle N \right\rangle<D$) are formed these are weakly affected by the spherical confinement. This is observed in the profiles of the average densities which shows relatively flat spatial distribution of filament segments.  In contrast the average density distribution of networks dominated by long linear filaments ($\left\langle N \right\rangle>D$) has a spatial organisation with a higher density of filaments in the periphery of the cell than in its centre. This shows that the latter type of networks are under strong confinement and the filaments align parallel to the cell wall in order to minimise the free energy of the system \cite{sakaue2007semiflexible,Azari2015,cifra2012weak,chen2006free,nikoubashman2017semiflexible}.  {We have obtained, though the network is strongly confined, that under certain conditions filaments near walls are perpendicular to the walls.}

  Looking at the density and orders profiles and by comparing the size of confining region and the persistence length to the contour length we classify the actin cytoskeletal networks inside confining spherical geometry according to their structure, their behaviour and properties in a phase diagram where there is no discontinuity numerically observable in the free energy of the system. Crossing the solid black line in these figures (cf.~Figures \ref{fig:Figp1} and \ref{fig:Figp2}) is, however, defined by a definite change of sign of the density curvature in the centre of the cell. Since the cells are not homogeneous with order parameter varying along
the radius, and the cell is finite, this is not a classical phase transition, but more like a cross-over between regimes of fundamentally different occupation of filaments inside the cell.

  The model we have developed can be reproduced for cells with various geometries and can be tested experimentally. 
   Fluorescent imaging of cells is able to inform one about filament density and it is also possible to gain orientational data (Tsugawa \etal~\cite{Tsugawa2016} showed this for microtubules).    
  This study, though it is a strongly simplified picture of a real cytoskeleton, can contribute to understand how branching cytoskeletal networks of filaments self-organise and align under the effect of confinement that the cell membrane introduce.
  The knowledge of how the structures of these networks relate to their physical properties and conformations is relevant for manufacturing of artificial living cells \cite{xu2016artificial} with different structures and geometries.


\begin{acknowledgement}
	This work is based on the research supported
	in part 
	by the National Research Foundation of South Africa (Grant No.~99116). 
	This work is based on funding of the SA-UK Newton Fund.
	SA is generously sponsored by the Organisation for Women in Science in the Developing World and SIDA.
\end{acknowledgement}


\appendix

\section{Density functional derived from the function $\psi$}
\label{App:DensityDerivation}

We implement the conventional definition for the density in a grand ensembles using eq.~\eqref{eq:basic-filament-density1} as applied to the partition function eq.~\eqref{eq:partition-fn-lin} or eq.~\eqref{eq:partfnpsi} with the more general $\psi$. Diagrammatically this entails the sum over all possible ``prunings'' at a $z$ of all possible trees. We note that $\psi$ is a functional of ${z}$, which leads to
\begin{subequations}
	\begin{align}
		\frac{\delta \, \mathfrak{Z}}{\delta \, z(\vv{r},\uv{n})} & = 
			\frac{1}{\mathfrak{Z}}\left[
			 	\psi(\vv{r},\uv{r}) + \int \dd^3 r' \, \dd^2 n' \frac{\delta \psi(\vv{r}'\,\uv{n}'}{\delta \, z(\vv{r},\uv{n})}
			\right] 
			\label{eq:funcderivA}
			\\
			\text{where } \frac{\delta \psi}{\delta\, z} & = 
	\int \,\, G\, w_0 \frac{\delta \psi}{\delta\, z} 
 \nonumber \\ &  +
 \int  \,\, G\, \xi \, \zeta\, \left(\frac{\delta \psi}{\delta\, z}\,\psi + \psi \frac{\delta \psi}{\delta\, z}\right).
 \label{eq:funcderivSC}
	\end{align}
\end{subequations}
Eq.~\eqref{eq:funcderivSC} can be expanded self-consistently. By noting the integration in eq.~\eqref{eq:funcderivA} one can then show that eq.~\eqref{eq:filamentdensityfirstinstance} and eq.~\eqref{eq:basic-filament-density1} emerge.

For the functional derivative of the grand partition function with respect to the junction fugacity $\zeta$ an analogous procedure can followed. Diagrammatically this is the sum of all ``prunings'' of trees precisely at all possible junctions. This pruning leads always to three trees emerging from the scission, two of which grow from a stem, and the remaining tree described from its leaf.

\section{Unconfined limit}
\label{app:unconfined}

Without any confinement whatsoever the centre of mass of the network is translationally and rotationally invariant. Quantities $\psi$ and $\tpsi$ must therefore not depend on the position and orientation vectors, $\vv{r}$ and $\uv{n}$, \textit{i.e.}~they are are constant.

Using the appropriate normalizations of Boltzmann weights, we see that the constant values of $\psi_0$ and $\tpsi_0$, corresponding to this unconfined limit, must satisfy
\begin{subequations}
	\begin{align}
	\psi_0 = & 1 + z_0 \psi_0 + \zeta_0 \psi^2 \text{ and}\label{eq:psi-unconfined}\\
	{\tpsi}_0 = & 1 + z_0 \tpsi_0 + \zeta_0 \psi \tpsi_0. \label{eq:tildepsi-unconfined}
\end{align}
\end{subequations}
Solutions only exist for values of $\zeta_0$ and $z_0$ with a positive discriminant in eq.~\eqref{eq:psi-unconfined}
\begin{equation}
	\left(1-z_0\right)^2 \geq 4 \zeta_0.
	\label{eq: condition}
\end{equation}
The single physical solution of the quadratic equation is 
\begin{subequations}
	\begin{align}
		\psi_0 = & \frac{1}{2\zeta_0}\left( 1-z_0 - \sqrt{(1-z_0)^2-4\zeta_0} \right) \text{ and}\\
		\tpsi_0 = & \psi_0. 
	\end{align}
\end{subequations}
With $n_\text{d}$ the number of discrete direction components per site, and $N$ the macroscopically large number of lattice sites, this leads to an expression for the density of filaments in unconfined conditions:
\begin{equation}
	\varrho_{0,\text{f}} =  \frac{n_\text{d} z_0 \psi_0 \left(1+ \psi_0\right)}{ 1+ n_\text{d} z_0 N\psi_0}.
\end{equation}
For actual junctions we have
\begin{equation}
	\varrho_{0,\text{b}} =  \frac{n_\text{d}^2 \zeta_0 \psi_0^3}{ 1+ n_\text{d} z_0 N\psi_0}.
\end{equation}
In the limit $N\rightarrow \infty$ this leads to the number of filaments and branching junctions as
\begin{subequations}
	\begin{align}
		N_{0,\text{f}} = & 1 + \psi_0 \text{ and}\\
		N_{0,\text{b}} = & \frac{\zeta_0}{z_0} n_\text{d}^2 \psi_0^2.
	\end{align}
\end{subequations}

\section{Computation technique}
\label{app:computation}

We describe the numerical method for solving the equations for $\psi$ and $\widetilde{\psi}$ for linear and  branched filament networks under confinement. The computation consists of solving the nonlinear integral equation \eqref{eq:psiIntegralEq} iteratively for the quantity $\psi(\pmb{r},\hat{\pmb{n}})$  together with the eq.~\eqref{eq:psitilde} for $\widetilde{\psi}(\pmb{r},\hat{\pmb{n}})$. 

We do this in three and two dimensions.
Generally plant cells have a very large vacuole which occupies most of the interior space of the cell and only a  thin area is occupied by the cytoskeleton leading to the two-dimensional shape of the cytoskeleton \cite{Gunning1996}; the cytoskeleton of animal cells presents a 3d shape \cite{Clark2013}. We choose to do the numerical implementations on 2d and 3d triangular lattices.
 
\subsection{Realisations of 2d and 3d lattices}
\subsubsection{2d triangular lattice}

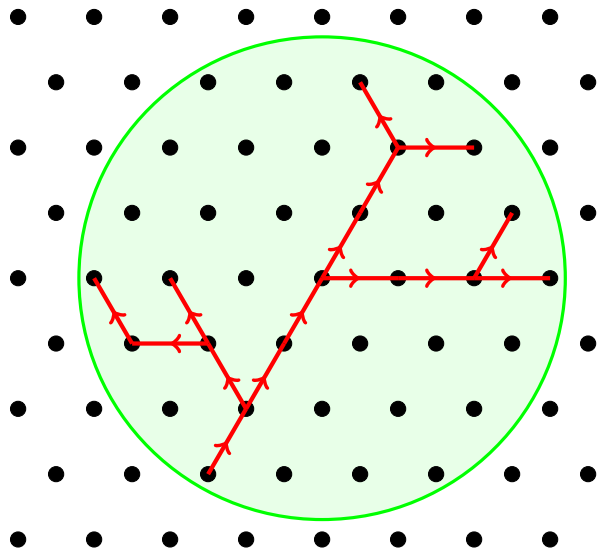
\begin{figure}[!h]
\begin{tikzpicture}[scale=1.0]
	\begin{scope}
		\clip (-0.2,-0.2) rectangle (7.2,7.2);
		\fill[green!30,opacity=0.3] (4,0.86603*4) circle (3.2);
	\end{scope}
	\draw[very thick, green] (4,0.86603*4) circle (3.2);
	\foreach \x in {0,...,7}
		\foreach \y in {0,...,3}
			{\draw[fill=black,black] (0.5+\x,2*0.86603*\y+0.86603) circle (0.1);};
	\foreach \x in {0,...,7}
		\foreach \y in {0,...,4}
			{\draw[fill=black,black] (\x,2*0.86603*\y) circle (0.1);};
	\begin{scope}[ultra thick,red,decoration={markings,mark=at position 0.5 with {\arrow{>}}}] 
		\draw[postaction={decorate}] (2.5,0.86603)--(3,2*0.86603);
		\draw[postaction={decorate}] (3,2*0.86603)--(3.5,3*0.86603);
		\draw[postaction={decorate}] (3.5,3*0.86603)--(4,4*0.86603);
		\draw[postaction={decorate}] (4.0,4*0.86603)--(4.5,5*0.86603);
		\draw[postaction={decorate}] (4.5,5*0.86603)--(5,6*0.86603);
		\draw[postaction={decorate}] (5,6*0.86603)--(4.5,7*0.86603);
		\draw[postaction={decorate}] (5,6*0.86603)--(6,6*0.86603);
		\draw[postaction={decorate}] (4,4*0.86603)--(5,4*0.86603);
		\draw[postaction={decorate}] (5,4*0.86603)--(6,4*0.86603);
		\draw[postaction={decorate}] (6,4*0.86603)--(7,4*0.86603);
		\draw[postaction={decorate}] (6,4*0.86603)--(6.5,5*0.86603);
		\draw[postaction={decorate}] (3,2*0.86603)--(2.5,3*0.86603);
		\draw[postaction={decorate}] (2.5,3*0.86603)--(2,4*0.86603);
		\draw[postaction={decorate}] (2.5,3*0.86603)--(1.5,3*0.86603);
		\draw[postaction={decorate}] (1.5,3*0.86603)--(1.0,4*0.86603);
   	\end{scope}
\end{tikzpicture}
\caption{An example of a configuration of branching network on a 2d triangular lattice. Black dots indicate the lattice and the network is confined to the region within the green circle. The chains or chain segments of the tree-like network are given in red, with arrows for the orientation of the segment.}
\label{fig:2dexample}
\end{figure}

We consider a 2d triangular lattice model of $N \times N$ sites. This lattice has the coordination number six. Fig.~\ref{fig:2dexample} shows one realisation of a branching network within a confined region on the triangular lattice.
A monomer starting at a site can point in the directions along the six adjoining bonds. Each bond is also associated with a fugacity $z$, that includes $z_0$, as well as the constraint (\textit{cf.}~eq.~\eqref{eqs:constraintimpl}. 
More importantly, the triangular lattice allows branching of filaments at multiples of $60^{\circ}$ angles, which is in the range of the angles of branching via the Arp2/3 protein complex, at about $70^{\circ}$~\cite{MalyBorisy2001,quint2011}.

The nearest neighbours are reached on the lattice using the basis unit vectors and  integer indices of the lattice points. In a 2d triangular lattice with unit bond length, the lattice indices of the nearest neighbours to the point $P = (X, Y)$ are $(X\pm 1,Y)$, $(X,Y\pm 1)$, $(X\mp 1,Y\pm 1)$.
$\uv{x}=\pm (1,0)$, $\uv{y}=\pm(\frac{1}{2},\frac{\sqrt{3}}{2})$, and $\uv{w}=\pm(\uv{y}-\uv{x})=\pm (-\frac{1}{2},\frac{\sqrt{3}}{2})$. This can be implemented in different ways, \textit{e.g.} 
\begin{equation}
\overrightarrow{OP}= (X_{g}, Y_{g})= X \uv{x}+Y\uv{y}.
\end{equation}
where $X$, $Y \in \mathbb{Z}$; or by considering alternating rows
\begin{equation}
	(X_{g}, Y_{g})= \begin{cases}
		(X ,\frac{\sqrt{3}}{2}Y), & Y  \text{ even}\\
(X+\frac{1}{2} ,\frac{\sqrt{3}}{2}Y), &   Y \text{ odd}
	\end{cases}.
\end{equation}

\subsubsection{3d triangular lattice}

We also consider the 3d triangular lattice with $N\times N\times N$ sites. Each lattice point has 12 nearest-neighbours. The network segments lie along the bonds, meaning that each monomer has 12 possible directions. The 3d triangular lattice consists of layers of square lattices in parallel planes and is formed by a closest cubic packing of spheres \cite{Gillespie2009}. To obtain a 3d triangular lattice, besides the two primary unit vectors $\uv{x}$ and $\uv{y}$  specifying the $x$ and $y$ axes of each square lattice, a primary axis along a unit vector $\uv{z}$ is added. This lies at an angle of $2\pi/3 $ from both $x$ and $y$.  The distance between two adjacent planes is $\sqrt{2}/2$. The planes support two consecutive layers of square lattices. We have the same distance between all nearest-neighbours. Again the choice of lattice maintains symmetry and allows us to approach the actual branching angle of actin filaments.

The twelve unit vectors  
are
\begin{equation}
\begin{cases} \uv{x}= \pm(1,0,0)\\ \uv{y}= \pm(0,1,0)\\ \uv{z}= \pm(-\frac{1}{2},-\frac{1}{2},\frac{\sqrt{2}}{2})\\ \uv{u}=\pm(\uv{x}+\uv{z})=\pm(\frac{1}{2},-\frac{1}{2},\frac{\sqrt{2}}{2})\\ \uv{v}=\pm(\uv{y}+\uv{z})=\pm(-\frac{1}{2},\frac{1}{2},\frac{\sqrt{2}}{2})\\ \uv{w}=\pm (\uv{x}+\uv{y}+\uv{z})=\pm(\frac{1}{2},\frac{1}{2},\frac{\sqrt{2}}{2})\end{cases}.
\end{equation}
Each lattice point position $P$ on the lattice is represented by integer coordinates $(X,Y,Z)$ where $X,Y,Z \in \mathbb{Z}$ with nearest neighbours of $P$:
\begin{multline}
\left\lbrace (X\pm 1,Y,Z),\quad (X,Y\pm 1,Z), \quad (X,Y,Z\pm 1),\right.\\ \left.(X\pm 1,Y,Z\pm 1), \quad (X,Y\pm 1,Z\pm 1), \right. \\ \left.   (X\pm 1,Y \pm 1,Z\pm 1)\right\rbrace \end{multline}
The corresponding Cartesian position vector of $P$ from the origin of the lattice $O$ is given by
\begin{equation}
\overrightarrow{OP}= (X_{c}, Y_{c},  Z_{c})= X \uv{x}+ Y\uv{y}+ Z\uv{z},
\end{equation}
meaning that
$X_{c}=  X-\frac{1}{2}$, 
$Y_{c}= Y-\frac{1}{2}$, and 
$Z_{c}= \frac{Z}{\sqrt{2}}$.

\subsubsection{Description and Algorithm}

Under the two chemical potential fields $z$ and $\zeta$ ($z$ associated to actin monomers and responsible for filaments elongation and $\zeta$ associated to Arp2/3 protein and responsible for branching), each monomer occupies a lattice bond. We show an example of network  configuration in Figure \ref{fig:2dexample}.
Our network bonds have polarisation.
We define a confining spherical region $\mathbb{L}$ that has a diameter $D$. The persistence length $\ell_{\text{p}}$ (via $w$) is chosen to be of the same order as $D$. 
	The additional length scale that will play a role is $N$ for the degree of polymerisation, which is largely determined by the fugacities.

The fugacities $z(\pmb{r},\hat{\pmb{n}})$ and $\zeta(\pmb{r},\hat{\pmb{n}}_{1},\hat{\pmb{n}}_{2})$ are associated   to filament elongation and branching controls, respectively, within the spherical region boundary limit. Explicitly the boundary conditions enter as follows
\begin{equation}
        z(\pmb{r}, \hat{\pmb{n}}) = z_{0} \times
        \begin{cases}
                1, & \text{if } \pmb{r} \in\mathbb{L} \text{ and } \pmb{r}+ \hat{\pmb{n}} \in \mathbb{L}\\
                0, & \text{otherwise}
        \end{cases}
        \label{eq:eq23}
\end{equation}
\begin{equation}
\zeta(\pmb{r},\hat{\pmb{n}}_{1},\hat{\pmb{n}}_{2})=\zeta_{0} \times
        \begin{cases}
                1, & \text{if } \pmb{r} \in\mathbb{L} \\ & \text{ and } \pmb{r}+ \hat{\pmb{n}}_{1} \in \mathbb{L} \\ & \text{ and } \pmb{r}+\hat{\pmb{n}}_{2} \in \mathbb{L} \\
                0, & \text{otherwise}
        \end{cases}
\end{equation}
We check if $\pmb{r}-\pmb{r}_{c}$ and $\pmb{r} + \hat{\pmb{n}}-\pmb{r}_{c}$ are positions inside the sphere for nonzero fugacity. Here $\pmb{r}_{c}$ is the position of the centre of the sphere.

We now can compute the nonlinear integral/sum equations \ref{eq:psi} and \ref{eq:psitilde} by an iteration procedure. We start with initial guesses $\psi_{0}$ and $\tilde{\psi}_{0}$ on the lattice, that are inserted into weighted versions of the right-hand sides of equations \eqref{eq:psiIntegralEq} and \eqref{eq:psitilde} repeatedly until the answers converge. 
 
Either we have zero occupancy, \textit{i.e.} there is no filament along a bond, and then  $\psi(\pmb{r},\hat{\pmb{n}})$ and $\tilde{\psi}(\pmb{r},\hat{\pmb{n}})$ are set to 1, or bonds are formed  and then $\psi(\pmb{r},\hat{\pmb{n}})$ and $\tilde{\psi}(\pmb{r},\hat{\pmb{n}})$ are locally calculated.
The filaments can grow from any point of the lattice as long as this monomer is inside the confining region. So a monomer at position $\pmb{r}$  connects with a monomers at position  $\pmb{r}_{1}=\pmb{r}+\ell \hat{\pmb{n}}$ where  $\ell=1$ is the lattice unit length. The unit vector $\hat{\pmb{n}}$ is one of the 6 (2d triangular lattice), 12 (3d triangular lattice). The monomer at position $\pmb{r}_{1}$  branches with two neighbours respectively at positions $\pmb{r}_{2}$ and $\pmb{r}_{3}$   on the lattice  and form a tree-like filament or as junction as follow:  
\begin{equation}
\rightarrow \pmb{r}_{1}
\begin{cases}
\nearrow \pmb{r}_{2}=\pmb{r}_{1}+ \hat{\pmb{n}}_{1}\\
\searrow \pmb{r}_{3}= \pmb{r}_{1}+ \hat{\pmb{n}}_{2}.
\end{cases}
\end{equation}
A new $\psi$ at starting at $\pmb{r}$ and with direction $\hat{\pmb{n}}$ is then computed iteratively. For 
example, the first iteration gives
\begin{align}
 \psi(\pmb{r},\hat{\pmb{n}})& = 1+ \sum_{\hat{\pmb{n}}_{1}}  w_{0}(\hat{\pmb{n}}, \hat{\pmb{n}}_{1})  z(\pmb{r},\hat{\pmb{n}}_{1})\psi_{0}(\pmb{r}_{1},\hat{\pmb{n}}_{1})\\
 & +\sum_{\hat{\pmb{n}}_{1}}\sum_{\hat{\pmb{n}}_{2}}\xi( \hat{\pmb{n}}, \hat{\pmb{n}}_{1}, \hat{\pmb{n}}_{2})  \zeta(\pmb{r},\hat{\pmb{n}}_{1},\hat{\pmb{n}}_{2})\psi_{0}(\pmb{r}_{2},\hat{\pmb{n}}_{1})\psi_{0}(\pmb{r}_{3},\hat{\pmb{n}}_{2}).
 \label{eq:eq34d}
\end{align}
This is repeated until the desired convergences is reached.

The computation of $\tilde{\psi}(\pmb{r},\hat{\pmb{n}})$ is similar to the computation of  $\psi(\pmb{r},\hat{\pmb{n}})$ with a few minor difference 
\begin{eqnarray}
\pmb{r}\rightarrow \pmb{r}_{1}&=&\pmb{r}- \hat{\pmb{n}}\\
\rightarrow \pmb{r}_{1}
\begin{cases}
\nearrow \pmb{r}_{2}=\pmb{r}_{1}+ \hat{\pmb{n}}_{1}\\
\searrow \pmb{r}_{3}= \pmb{r}_{1}-\hat{\pmb{n}}_{2}.
\end{cases}
\end{eqnarray}

We then compute the grand partition function, the spatial density distributions and the nematic order field for different type of networks, typed according to their stiffness and their structures, since these all quantities depend on $\psi$ and $\widetilde{\psi}$.

\section{Persistence length}
\label{app:persistence}

Since the chains exist on a lattice with bonds connecting only nearest neighbours, the semi-flexible chain can be mapped onto a Potts model.
We wish to determine  the persistence length $\ell_{\textbf{p}}$ of the unconfined actin filaments as function the filament  bending modulus $\epsilon$. We compare $\ell_\text{p}$ with the size of the confining region. In our model the bending energy (with modulus) this is encoded in the Boltzmann weight $w$.

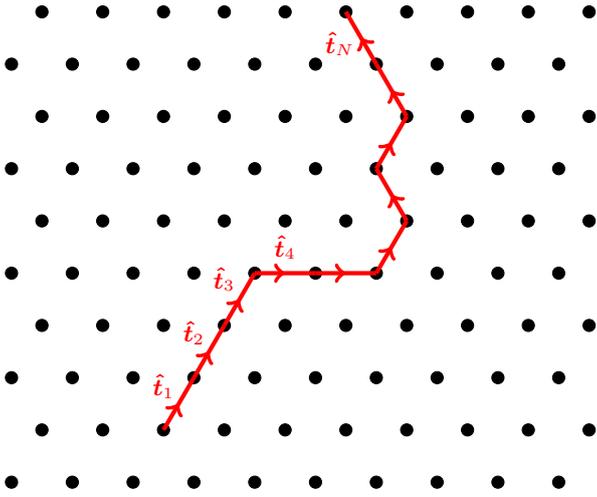
\begin{figure}[!h]
\begin{tikzpicture}[scale=0.8]
	\foreach \x in {0,...,9}
		\foreach \y in {0,...,4}
			{\draw[fill=black,black] (0.5+\x,2*0.86603*\y+0.86603) circle (0.1);};
	\foreach \x in {0,...,9}
		\foreach \y in {0,...,4}
			{\draw[fill=black,black] (\x,2*0.86603*\y) circle (0.1);};
	\begin{scope}[ultra thick,red,decoration={markings,mark=at position 0.5 with {\arrow{>}}}] 
		\draw[postaction={decorate}] (2.5,0.86603)--(3,2*0.86603);
		\node at (2.5, 1.8*0.88603) {$\uv{t}_1$};
		\draw[postaction={decorate}] (3,2*0.86603)--(3.5,3*0.86603);
		\node at (3, 2.8*0.88603) {$\uv{t}_2$};
		\draw[postaction={decorate}] (3.5,3*0.86603)--(4,4*0.86603);
		\node at (3.5, 3.8*0.88603) {$\uv{t}_3$};
		\draw[postaction={decorate}] (4,4*0.86603)--(5,4*0.86603);
		\node at (4.5, 4.4*0.88603) {$\uv{t}_4$};
		\draw[postaction={decorate}] (5,4*0.86603)--(6,4*0.86603);
		\draw[postaction={decorate}] (6,4*0.86603)--(6.5,5*0.86603);
		\draw[postaction={decorate}] (6.5,5*0.86603)--(6,6*0.86603);
		\draw[postaction={decorate}] (6,6*0.86603)--(6.5,7*0.86603);
		\draw[postaction={decorate}] (6.5,7*0.86603)--(6,8*0.86603);
		\draw[postaction={decorate}] (6,8*0.86603)--(5.5,9*0.86603);
		\node at (5.4, 8.2*0.88603) {$\uv{t}_N$};
   	\end{scope}
\end{tikzpicture}
\caption{A configuration of a chain on the hexagonal lattice, with $\uv{t}_i$ indicating the unit vector orientation of the the $i$th bond in the chain.}
\label{fig:Fig2}
\end{figure}

The persistence length is determined from the exponential decay of the tangent-tangent correlation function of a semi-flexible walk, \textit{cf.} Fig.~\ref{fig:Fig2}. 
We calculate the correlation  $\langle \pmb{t}_{i}\cdot\pmb{t}_{N+i}\rangle$. The Boltzmann weight associated to the bending energy between adjacent bonds is
$w(i,i+1)=  \langle \pmb{t}_{i}|\mathcal{T}|\pmb{t}_{i+1}\rangle$
where the $\pmb{t}_{i}$ are tangent vectors describing all possible orientations of the bonds of the chain on a defined lattice, expressed in a suitable basis for the transfer matrix $\mathcal{T}$.
To calculate $\langle \pmb{t}_{a}\cdot\pmb{t}_{N+a}\rangle$ we compute partition function $\mathcal{Z}$ of the full polymer chain of $M=a+b+N$ monomers:
\begin{equation}
\mathcal{Z}=\sum_{\pmb{t}_{i}}\prod_{i=1}^{M-1}w(i,i+1).
\end{equation}
The correlation function between the tangent vector $\pmb{t}_{a}$ and $\pmb{t}_{N+a}$ is
\begin{equation}
\langle \pmb{t}_{a}\cdot\pmb{t}_{N+a}\rangle=\frac{1}{\mathcal{Z}}\sum_{\pmb{t}_{i}}  \pmb{t}_{a}\cdot\pmb{t}_{N+a}\prod_{i=1}^{M-1}w(i,i+1).
\end{equation}
Using the following relation
\begin{equation}
\langle \pmb{t}_{a}\cdot\pmb{t}_{N+a}\rangle \sim e^{-{N}/{\ell_{\textbf{p}}}}, \text{ for }N\gg 1,
\end{equation}
we compute the persistence length $\ellp$ of the polymer chain 
\begin{equation}
\ell_{\text{p}}\sim -\frac{N}{\ln{(\langle \pmb{t}_{a}\cdot\pmb{t}_{N+a}\rangle )}}.
\end{equation} 
We choose the ratio of the persistence length to the size of the confining cell at a value greater or equal to 1. This choice makes the filament semi-flexible on the scale of the cell. The degree of polymerisation of the filament $\left\langle N \right\rangle$ is also relevant. We investigate difference scenarios:
\begin{itemize}
\item $\ell_{\text{p}}\simeq \left\langle N \right\rangle >D$ --- semi-flexible polymer chain that is subject to strong effect of the confinement;
\item $\ell_{\text{p}}\simeq \left\langle N \right\rangle \sim D$, --- confinement, length, and persistence length of comparable length scale; and
\item $\ell_{\text{p}}\simeq \left\langle N \right\rangle < D$ --- weak confinement effect.
\end{itemize}

\begin{figure}[!h]
\includegraphics[width=0.4\textwidth]{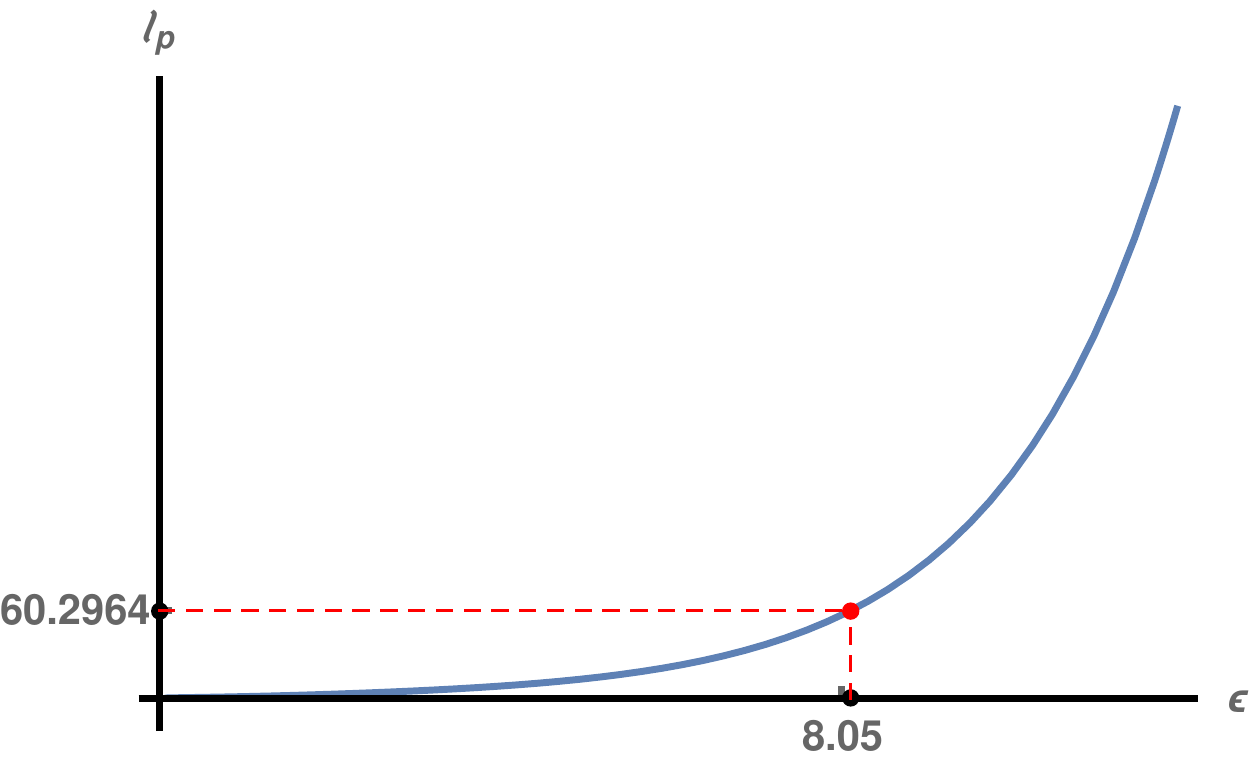}
\caption{{Plot of the persistence length of a polymer chain living on the 2d triangular lattice as function of} $\epsilon$.}
\label{fig:Fig3}
\end{figure}

 {Figure \ref{fig:Fig3} shows an example of the relationship between $\epsilon$ and $\ell_\text{p}$.
For our numerical calculations we choose a value for the bending rigidity $\epsilon$ giving a persistence $\ell_{\text{p}}$ such that the ratio $\ell_\text{p}/D$ stays fixed to $1.2$. 
For the 2d triangular lattice $\ell_{\text{p}}=60.2964$ lattice bond lengths, with taken to be $D=50$ bond lengths, \textit{i.e.} ${\ell_{\text{p}}}/{D}=1.2 $. For the 3d triangular lattice our choice of $\epsilon$ is such that $\ell_\text{p}=30.0077$ with $D=25$.}


\bibliographystyle{epj}

\bibliography{bibliography}
	
\end{document}